%% file: main.tex
\documentclass[12pt]{article}
\RequirePackage[l2tabu, orthodox]{nag}
\usepackage[main=english]{babel}
\usepackage[rm={lining,tabular},sf={lining,tabular},tt={lining,tabular,monowidth}]{cfr-lm}
\usepackage[T1]{fontenc}
\usepackage[utf8]{inputenc}
\usepackage[pdftex]{graphicx}
\usepackage{amsthm,amssymb,latexsym,gensymb,mathtools,mathrsfs}
\usepackage{epstopdf,enumitem,microtype,dcolumn,booktabs,url,fancyhdr}
\usepackage{algorithm}
\usepackage{algpseudocode}
\usepackage{natbib}
\usepackage{bbm}
\usepackage[affil-it]{authblk}
\usepackage{fancyvrb}
\usepackage{caption, subcaption}
\usepackage{array}
\usepackage{booktabs}
\usepackage{pgfplots}
\usepackage{xinttools}
\usepgfplotslibrary{fillbetween}
\pgfplotsset{compat=1.16}
\usepackage{tikz}
\usepackage{multirow}
\usepackage[most]{tcolorbox}
\usepackage[hypertexnames=false]{hyperref}
\usepackage{cleveref}

\input{macros_general.tex}

\input{macros.tex}

\setlist{topsep=1ex,parsep=1ex,itemsep=0ex}
\setlist[1]{leftmargin=\parindent}
\setlist[enumerate,1]{label=\arabic*.,ref=\arabic*}
\setlist[enumerate,2]{label=(\alph*),ref=(\alph*)}

\graphicspath{.}

\newtheorem{prop}{Proposition}

\newtheorem{rmk}{Remark}

\newtheorem{ex}{Example}

\crefname{subsubsection}{Section}{Sections}
\Crefname{subsubsection}{Section}{Sections}
\crefname{ex}{Example}{Examples}
\Crefname{ex}{Example}{Examples}

\title{Propagating Surrogate Uncertainty in Bayesian Inverse Problems}
\author[1]{Andrew Gerard Roberts\thanks{Emails: \texttt{arober@bu.edu}, \texttt{dietze@bu.edu}, \texttt{huggins@bu.edu}}}
\author[2]{Michael Dietze}
\author[1,3]{Jonathan H. Huggins}

\affil[1]{Faculty of Computing and Data Sciences, Boston University}
\affil[2]{Department of Earth and Environment, Boston University}
\affil[3]{Department of Mathematics and Statistics, Boston University}
\date{}

\begin{document}
\maketitle

\begin{abstract}
Standard Bayesian inference schemes are infeasible for inverse problems 
with computationally expensive forward models. A common solution is to 
replace the model with a cheaper surrogate. To avoid overconfident 
conclusions, it is essential to acknowledge the surrogate approximation by propagating 
its uncertainty. At present, a variety of distinct uncertainty propagation methods 
have been suggested, with little understanding of how they vary.
To fill this gap, we propose a mixture distribution termed the \textit{expected posterior (EP)} 
as a general baseline for uncertainty-aware posterior approximation, justified by decision theoretic 
and modular Bayesian inference arguments. We then investigate the 
\textit{expected unnormalized posterior (EUP)}, a popular heuristic alternative, analyzing when 
it may deviate from the EP baseline. Our results show that this heuristic can break down when 
the surrogate uncertainty is highly non-uniform over the design space, as can be the case when
the log-likelihood is emulated by a Gaussian process. Finally, we present the
\textit{random kernel preconditioned Crank-Nicolson (RKpCN)} algorithm, an approximate 
Markov chain Monte Carlo scheme that provides a practical EP approximation in the challenging
setting involving infinite-dimensional Gaussian process surrogates.
\end{abstract}

\section{Introduction}
Simulation-based computer models are key tools for studying complex systems within 
the physical, biological, and engineering sciences. Such models often have 
unknown parameters that must be estimated (i.e., calibrated) using observational data.
Quantifying the uncertainty in these estimated values is crucial for downstream 
decision making. While Bayesian methods are particularly well-suited to this task, 
standard Bayesian inference algorithms such as Markov chain Monte Carlo (MCMC) 
are hindered by the computational cost of many simulation models. A popular solution 
is to use a small set of expensive simulations to train a statistical approximation 
of the simulator \citep{gramacy2020surrogates}. 
This surrogate (i.e., emulator) is then used as a drop-in replacement 
for the true computer model, enabling the application of algorithms like MCMC. 
This modular surrogate-based Bayesian workflow has seen widespread use across
a variety of applications \citep{FerEmulation,FadikarAgentBased,idealizedGCM,
trainDynamics,FATES_CES,CLMBayesianCalibration}.

Despite significant advances in surrogate modeling, fitting a 
highly accurate emulator under a limited computational budget is typically impossible,
invariably implying the presence of errors in the surrogate-based posterior approximation.
Ignoring these errors can lead to biased results 
with miscalibrated uncertainties \citep{BilionisBayesSurrogates,BurknerSurrogate}.
It is thus crucial to acknowledge and propagate this additional source of uncertainty in 
surrogate-based Bayesian workflows. 
Probabilistic surrogates such as Gaussian processes (GPs; \citet{gpML,gramacy2020surrogates}) 
and probabilistic neural networks \citep{deepEnsembles,BayesOptNN} provide a notion of 
predictive uncertainty that be can utilized to this end.

While in principle surrogate and calibration parameters can be learned 
jointly (e.g., \citet{KOH}), in practice it is more common to conduct inference 
for these quantities in two distinct stages \citep{modularization,PlummerCut}.
Such decoupling has several benefits, avoiding computation for the larger joint
model and preventing a misspecified calibration likelihood from affecting
inference for the surrogate parameters \citep{modularization}.
 However, it leaves open the 
question as to the ``correct'' approach for propagating surrogate uncertainty 
within the posterior approximation in the second stage.
A variety of uncertainty-aware posterior 
approximations have been proposed, but little guidance exists on choosing a particular 
method \citep{BilionisBayesSurrogates,StuartTeck1,VehtariParallelGP,
BurknerSurrogate,BurknerTwoStep,FerEmulation}. Moreover, previous studies have 
explicitly cited computational challenges as a key factor in determining their 
approach \citep{VehtariParallelGP,StuartTeck2}. In this paper, we start by setting 
aside the computational considerations in order to identify a theoretically-justified 
baseline posterior approximation termed the \textit{expected posterior (EP)}. In general, 
EP-based inference is computationally demanding and difficult to implement for 
infinite-dimensional surrogates such as GPs. To alleviate these 
challenges, one can consider either (1) abandoning the EP in favor of convenient heuristic
alternatives, or (2) developing approximate inference methods that directly target the EP.
We analyze the \textit{expected unnormalized posterior (EUP)}, a commonly-used heuristic
falling under the former category. While the EUP approximation is often reasonable, we 
demonstrate that it can be unstable when the uncertainty in the surrogate-induced approximation 
of the posterior density is highly variable over the parameter space. We then show that the latter
approximate computation viewpoint can be fruitful in achieving a more robust EP approximation.
In particular, we present \textit{random kernel preconditioned Crank-Nicolson (RKpCN)}, an
approximate MCMC algorithm targeting the EP that is well-defined for GP surrogates.

The remainder of this paper is structured as follows.
\Cref{sec:surrogates-intro} introduces the modular surrogate-based Bayesian workflow.
In \Cref{sec:EP} we derive the EP as a Bayes' estimator, and discuss connections 
with the so-called cut posterior distribution. In \Cref{sec:approximating-ep}
we analyze the EUP as an EP approximation and highlight practical 
takeaways for common applications in which GPs are used to emulate 
forward models or log-densities.
\Cref{sec:computation} presents an approximate MCMC scheme directly 
targeting the EP, and describes connections with alternative inference algorithms. 
\Cref{sec:experiments} contains numerical experiments, and 
\Cref{sec:conclusion} concludes. Proofs, derivations, and technical details
are given in the appendix.

\section{Surrogates for Bayesian Inference} \label{sec:surrogates-intro}
We begin by introducing the Bayesian inference setting, including the challenges
associated with Bayesian inverse problems involving expensive forward models. 
We then describe the common two-stage surrogate modeling pipeline, and highlight 
several different strategies for integrating surrogates within a Bayesian analysis.

\subsection{Bayesian Inference Setting}
We consider the general goal of estimating parameters $\Par \in \parSpace \subseteq \R^{\dimPar}$ given 
observations $\obs \in \obsSpace \subseteq \R^{\dimObs}$ within a Bayesian framework.
A Bayesian model consists of a joint probability distribution $p(\Par, \obs)$, defined by 
specifying a prior density $\priorDens(\Par)$ and likelihood function $\lik(\Par; \obs)$.
The goal is then to summarize the posterior distribution 
\begin{align}
&\postDensNorm(\Par) \Def p(\Par \given \obs) = \frac{1}{\normCst} \priorDens(\Par) \lik(\Par; \obs), 
&&\normCst = \int_{\parSpace} \priorDens(\Par) \lik(\Par; \obs) d\Par. \label{eq:post_dens_generic}
\end{align}
While closed-form characterizations are typically thwarted by the intractable normalizing constant
$\normCst$, approximate posterior samples can be simulated using MCMC algorithms, which only require 
access to pointwise evaluations of the unnormalized density 
$\postDens(\Par) \Def \priorDens(\Par) \lik(\Par; \obs)$.
However, such methods commonly require $10^5 - 10^7$ iterations, with each iteration 
involving at least one query to the density $\postDens(\Par)$.
In various engineering and scientific applications, computing $\lik(\Par; \obs)$ (and thus
$\postDens(\Par)$) requires running an expensive computer simulation. 
This renders MCMC infeasible, motivating the need for inference schemes
that use only a small set of evaluations of $\postDens(\Par)$.
 
\subsection{Bayesian Inverse Problems} \label{sec:bip}
The challenge posed by computationally expensive density evaluations $\postDens(\Par)$ commonly 
arises in the Bayesian approach to inverse problems \citep{Stuart_BIP}. In this setting, 
the likelihood often takes the form $\obs = \fwd(\Par) + \noise$ for some forward model
$\fwd: \parSpace \to \obsSpace$. For a concrete example, we consider the problem of estimating the 
parameters in a system of ordinary differential equations (ODEs)
\begin{align}
\frac{d}{d\Time} \state(\Time, \Par) &= \odeRHS(\state(\Time, \Par), \Par), &&x(\timeStart, \Par) = \stateIC, \label{ode_ivp}
\end{align}
where the dynamics depend on parameters $\Par$. Each value for $\Par$ implies a different solution trajectory
$[\state(\Time, \Par)]_{\timeStart \leq \Time \leq \timeEnd}$, which we encode by the
map $\solutionOp: \Par \mapsto [\state(\Time, \Par)]_{\timeStart \leq \Time \leq \timeEnd.}$. The goal is then 
to identify the parameters that yield trajectories in agreement with observed data 
$\obs$, which is assumed to be some noise-corrupted function $\obsOp$ of the true trajectory. Thus, the 
likelihood is of the form 
\begin{align}
&\obs = \fwd(\Par) + \noise, &&\fwd \Def \obsOp \circ \solutionOp. \label{eq:additive-noise}
\end{align}
In practice, the ODE is solved numerically so $\solutionOp$ represents the map induced by a numerical 
solver. Therefore, in this setting the computational cost of computing $\postDens(\Par)$ stems from the 
dependence of the likelihood on $\fwd(\Par)$, and in particular on the solver $\solutionOp(\Par)$.

\subsection{Surrogate Targets for Bayesian Inference} \label{sec:surrogates-Bayes}
Given the cost of computing $\postDens(\Par)$, we seek to approximate
the posterior using a small set of queries to the posterior density. 
The surrogate modeling approach to this problem consists of constructing 
a regression-based approximation of the density $\postDens$. This approximate density is typically 
induced indirectly by the approximation of some underlying quantity 
on which $\postDens(\Par)$ depends. To make this explicit, let 
$\target: \parSpace \to \targetRange$ be the underlying map targeted for emulation. 
A regression model $\targetEm$ is fit to a set of exact simulator runs 
$\{(\Par_n, \target(\Par_n))\}_{n=1}^{\Ndesign}$, such that $\targetEm(\Par)$ provides at 
prediction of $\target(\Par)$. \Cref{ex:fwd-em,ex:ldens-em} describe the common strategies 
where $\target$ is chosen to be the forward model or log-likelihood, respectively.

In order to quantify the uncertainty introduced via this approximation, we consider 
emulators that provide predictions in the form of a probability distribution; 
i.e., $\targetEm(\Par)$ is a random vector and $\targetEm$ is a random function.
Let $\emDist(\cdot \given \Par)$ and $\emDist$ denote the distributions of 
$\targetEm(\Par)$ and $\targetEm$, respectively, and $\emE$ the expectation with respect to $\emDist$.
The randomness in $\targetEm$ typically quantifies the epistemic (reducible) uncertainty
due to the limited computational budget \citep{epistemicAleatoric}. 
Common models that provide predictive distributions include
Gaussian processes (GPs; \citep{gramacy2020surrogates}), 
Bayesian neural networks \citep{BayesOptNN}, deep ensembles \citep{deepEnsembles}, 
and Bayesian additive regression trees \citep{BARTReview}.

We consider the common surrogate modeling workflow in which $\targetEm$ is fit only to data generated by 
the simulator, and then substituted for $\target$ to approximate the original Bayesian model. This two-stage 
\textit{modular} pipeline (as opposed to fitting a joint Bayesian model) is practically convenient, and 
has been shown to outperform the joint Bayesian approach in the presence of likelihood misspecification
\citep{modularization, PlummerCut}. Using $\targetEm$ as a drop-in replacement for 
$\target$ induces predictive distributions over $\postDens$ and $\normCst$. To emphasize the dependence
of these quantities on $\target$, we write $\postDens(\Par; \target)$ and $\normCst(\target)$ and assume 
throughout that  $\postDens(\Par; \target)$ is dependent on $\Par$ only through $\target(\Par)$.
In the common case that the target appears only through the likelihood, we similarly write $\lik(\Par; \target, \obs)$.
The surrogate-induced approximations $\postDens(\cdot; \targetEm)$, $\normCst(\targetEm)$, and 
$\lik(\Par; \targetEm, \obs)$ are random quantities, with distributions given by the pushforward of $\emDist$ through
the respective maps. The following examples highlight two broad classes of surrogate targets commonly used
in the literature. A similar categorization is explored in \citet{StuartTeck1,StuartTeck2,GP_PDE_priors}. 

\begin{ex}[Forward Model Surrogate]
\label{ex:fwd-em}
In the Bayesian inverse problem setting from \Cref{sec:bip},
a natural approach is to target the underlying forward model 
$\Par \mapsto \fwd(\Par)$ (i.e., choose $\target \Def \fwd$), a strategy 
we refer to as \textbf{forward model emulation}.
This method consists of fitting a surrogate $\targetEm$ to the design 
$\{(\Par_n, \fwd(\Par_n))\}_{n=1}^{\Ndesign}$ and then using $\targetEm$
in place of $\fwd$. Much previous work has considered this strategy 
in the context of the additive noise model in \Cref{eq:additive-noise},
under the Gaussian noise assumption $\noise \sim \Gaussian(0, \likPar)$
\citep{StuartTeck1,GP_PDE_priors,hydrologicalModel,hydrologicalModel2,
Surer2023sequential,VillaniAdaptiveGP,weightedIVAR,idealizedGCM,CES}.
In this special case, the induced (unnormalized) posterior density approximation takes the form
\begin{align}
\postDens(\Par; \targetEm) &= \priorDens(\Par) \Gaussian(\obs \given \targetEm(\Par), \likPar). \label{eq:post-em-fwd-Gaussian}
\end{align}
\end{ex}

\begin{ex}[Log-Density Surrogate]
\label{ex:ldens-em}
Another popular strategy is to choose $\target$ as the map induced by the log-likelihood
$\Par \mapsto \log \lik(\Par; \obs)$
\citep{VehtariParallelGP,FATES_CES,trainDynamics,quantileApprox,
ActiveLearningMCMC,FerEmulation,StuartTeck1,random_fwd_models,
GP_PDE_priors,OakleyllikEm,JosephMinEnergy,AlawiehIterativeGP,gpEmMCMC}
or (unnormalized) log-posterior
$\Par \mapsto \log \left\{\priorDens(\Par)\lik(\Par; \obs)\right\}$
\citep{emPostDens,Kandasamy_2017,llikRBF,gp_surrogates_random_exploration,
landslideCalibration}.
We collectively refer to these strategies as \textbf{log-density emulation}. In the 
log-likelihood case, an emulator $\targetEm$ is fit to a design 
$\{(\Par_n, \log \lik(\Par_n; \obs)\}_{n=1}^{\Ndesign}$
and induces an unnormalized posterior density surrogate 
\begin{align}
\postDens(\Par; \targetEm) &= \priorDens(\Par) \Exp{\targetEm(\Par)}. \label{eq:post-em-llik}
\end{align}
The log-posterior case is similar, except that the effect of the prior is also 
approximated by the emulator, so the induced unnormalized posterior surrogate takes 
the form $\postEm(\Par) = \Exp{\targetEm(\Par)}$.
\end{ex}

\section{The Expected Posterior} \label{sec:EP}
The second stage in the modular surrogate workflow consists of using the 
trained emulator $\targetEm$ to approximate the posterior $\postDensNorm$.
A simple approximation may be constructed by plugging in the surrogate mean 
$\Em{\bar{\targetTraj}} \Def \emE[\targetEm]$
\begin{equation}
\postApproxNormMean(\Par) \Def \postDens(\Par; \Em{\bar{\targetTraj}}) / \normCst(\Em{\bar{\targetTraj}}),
\label{eq:mean-approx}
\end{equation}
but this ignores the emulator uncertainty, resulting in overconfident 
posterior inference. This raises the question of defining a posterior approximation 
that correctly propagates the uncertainty in $\targetEm$.
Given the lack of a unifying probabilistic model across the two inference stages, 
proper uncertainty quantification is not automatically given by standard Bayesian 
conditioning. Consequently, various uncertainty propagation methods have been 
proposed, each resulting in different posterior inferences 
\citep{BilionisBayesSurrogates,StuartTeck1,VehtariParallelGP,BurknerSurrogate,
FerEmulation}.
We identify and justify a mixture distribution named the 
\textit{expected posterior (EP)} as the correct distribution to target in 
modular surrogate-based inference. 

\subsection{Decision Theoretic Derivation} \label{sec:decision-theoretic}
Irrespective of the underlying target $\target$, the probabilistic emulator $\targetEm$ 
induces a random approximation of the posterior defined by plugging $\targetEm$
in place of $\target$; this yields
\begin{align}
\postDensNorm(\Par; \targetEm) 
&= \frac{\postDens(\Par; \targetEm)}{\normCst(\targetEm)},
&&\normCst(\targetEm) \Def \int_{\parSpace} \postDens(\Par; \targetEm) \d\Par, \label{eq:random-post}
\end{align}
which is referred to as the ``sample approximation'' in \citet{StuartTeck1}. For brevity, we write
 $\postEm(\cdot) \Def \postDens(\cdot; \targetEm)$, 
$\postNormEm(\cdot) \Def \postDensNorm(\cdot; \targetEm)$, and 
$\normCstEm \Def \normCst(\targetEm)$ when explicit reference to the underlying 
emulator is not necessary. The challenge of uncertainty propagation can be viewed as 
that of constructing a deterministic probability distribution that summarizes the uncertainty 
encoded in $\postNormEm$. To identify such a distribution, we adopt a Bayesian decision 
theoretic viewpoint and select an optimal distribution from a set of candidates $\qSpace$ that 
minimizes an expected loss $\emE[\loss(\postNormEm, q)]$. In other words, we seek a Bayes'
estimator 
\begin{equation}
\qDensOpt \in \argmin_{\qDens \in \qSpace} \emE[\loss(\postNormEm, q)]
\label{eq:variational-opt}
\end{equation}
with respect to a particular loss $\loss$ and space of densities $\qSpace$ over $\parSpace$. 
The following result provides the unique minimizer $\qDensOpt$ with respect to two common losses.

\begin{prop} \label{prop:EP-variational}
If the loss $\loss(\postNormEm, q)$ is chosen as the forward Kullback-Leibler (KL) divergence 
$\KL{\postNormEm}{q}$ or squared $L_2$ error $\norm{\postNormEm - q}_{L_2(\parSpace)}^2$,
then the optimization problem in \Cref{eq:variational-opt} is solved uniquely by 
\begin{equation}
\qDensOpt(\Par) = 
\emE \left[\postNormEm(\Par) \right]
= \int \postDensNorm(\Par; \targetTraj) \emDist(d\targetTraj).
\label{eq:ep-approx}
\end{equation}
\end{prop}

We thus take $\qDensOpt$ as the baseline for surrogate-based uncertainty propagation. 
This distribution has been considered previously in various 
contexts, but is not widely used in the surrogate modeling literature
\citep{trainDynamics,BurknerSurrogate,garegnani2021NoisyMCMC}.
This is likely due in part to computational difficulties, which we address in \Cref{sec:computation}.
Following \citet{BurknerSurrogate}, we refer to $\qDensOpt$ as the \textit{expected posterior (EP)},
denoted by $\postApproxEP \Def \qDensOpt$.

\subsection{Hierarchical Formulation}
The EP arises as the marginal of the 
joint distribution $\emDist(d\targetTraj) \postDensNorm(\Par; \targetTraj) d\Par$, which 
can be understood via the hierarchical model 
\begin{align}
&\targetTraj  \sim \emDist, 
&&\Par \given \targetTraj \sim \postDensNorm(\d\Par; \targetTraj).
\label{eq:ep-prob-model}
\end{align}
This perspective highlights the interpretation of the EP as a $\emDist$-weighted mixture of posteriors 
$\postDensNorm(\Par; \targetTraj)$, each induced by a particular emulator realization $\targetTraj$.
We assume throughout that $\targetEm$ is constructed such that 
trajectories of $\postNormEm$ are almost surely integrable, implying the sampling procedure
in \Cref{eq:ep-prob-model} is well-defined. See 
\citet{StuartTeck1,StuartTeck2,random_fwd_models,garegnani2021NoisyMCMC} for
technical conditions.

The EP also admits a marginal likelihood interpretation under the hierarchical model
\begin{equation}
 \targetTraj  \sim \emDist, \qquad \Par \sim \priorDens, \qquad 
\obs \given \targetTraj, \Par \sim \lik(\Par; \targetTraj, \d\obs)/\normCst(\targetTraj),
\end{equation}
yielding the equivalent characterization
\footnote{For clarity of exposition, we assume here that the prior $\priorDens$ is not part of the surrogate model.
A slight modification yields the general case, which encompasses log-posterior emulators.}
\begin{align}
&\postApproxEP(\Par) \propto \priorDens(\Par) \likApproxEP(\Par; \obs),
&&\likApproxEP(\Par; \obs) \Def \int \frac{\lik(\Par; \targetTraj, \obs)}{\normCst(\targetTraj)} \emDist(\d\targetTraj). 
\label{eq:lik-approx-EP}
\end{align}
Observe that the exact likelihood is replaced with an approximation averaged over $\targetTraj$ and 
weighted both by the surrogate predictive distribution and the marginal likelihood $\normCst(\targetTraj)$.

\subsection{Cut Posterior} \label{sec:cut}
In this section, we consider the setting where the two-stage surrogate workflow 
arises as an approximation to a coherent joint Bayesian model, in which case
 the EP can be viewed as a so-called cut posterior distribution
\citep{PlummerCut,cutInference,moduleModels,cutVar,cutVar2}.
In particular, suppose that the surrogate model is defined by specifying 
a prior $\targetTraj \sim \emDistPrior$ and likelihood $\emLik(\targetTraj; \emObs)$,
where $\emObs \Def \{(\Par_n, \target(\Par_n))\}_{n=1}^{\Ndesign}$ denotes
the emulator training data. The common example of a conjugate GP model 
corresponds to $\emDistPrior = \GP(\gpMean[0], \gpKer[0])$ and 
$\emLik(\targetTraj; \emObs) = \Gaussian(\target(\Par_{1:\Ndesign}) \given \targetTraj(\Par_{1:\Ndesign}), \tau^2 I)$.
This setup encompasses standard parametric Bayesian models as well.
We can thus consider the joint Bayesian model 
\begin{equation}
\jointKOH(\d\Par, \d\targetTraj, \d\obs, \d\emObs) \Def
\priorDens(\Par) \lik(\Par; \targetTraj, \obs) \emLik(\targetTraj; \emObs) 
\emDistPrior(\d\targetTraj) \, \d\Par \, \d\obs \, \d\emObs
\label{eq:koh-joint}
\end{equation}
over all unknowns $(\targetTraj, \Par)$. This fully Bayesian (non-modular) model 
is akin to the framework proposed in the seminal work of \citet{KOH}. Unfortunately, the joint model 
can produce counterintuitive results when $\lik(\Par; \target, \obs)$ is misspecified, stemming from 
the fact that both the simulated $\emObs$ and observational data $\obs$ inform inference
for the emulator \citep{modularization}. The following result shows that 
the EP can be viewed as an optimal approximation to the fully Bayesian model, 
subject to the constraint that $\obs$ is not allowed to inform $\targetTraj$.

\begin{prop} \label{prop:kl-cut-op}
Let $\postKOH$ denote the distribution of $(\Par, \targetTraj)$
given $(\obs, \emObs)$ under the joint $\jointKOH$. Also, let $\emDist$ denote 
the distribution of $\targetTraj$ given $\emObs$ under the joint 
$\emDistPrior(\d\targetTraj) \emLik(\targetTraj; \emObs) \d\emObs$. Then, 
\begin{equation}
\emDist(\d\targetTraj) \postDensNorm(\Par; \targetTraj)\d\Par
= \argmin_{\qMeas \in \qSpaceCut} \KL{\qMeas}{\postKOH},
\label{eq:kl-cut-opt}
\end{equation}
where 
\begin{equation}
\qSpaceCut \Def 
\left\{\qMeas(\d\Par, \d\targetTraj) : \int \qMeas(\d\Par, \cdot) = \emDist(\cdot) \right\}.
\label{eq:cut-dists}
\end{equation}
\end{prop} 
The optimum in \Cref{eq:kl-cut-opt} is precisely the joint distribution noted in the previous 
section, and also corresponds to the cut posterior with respect to the fully Bayesian model.
This gives a second variational justification for the EP, complementing the result in 
\Cref{prop:EP-variational}.

\paragraph{Related work.} In the surrogate modeling literature the EP has been considered 
in \citet{trainDynamics,BurknerSurrogate,garegnani2021NoisyMCMC}. In 
\citet{StuartTeck2,VehtariParallelGP}, the EP is briefly noted but deemed computationally impractical. 
This difference of opinion can be explained by the fact that these latter two papers are focused 
on GP surrogates, which present additional challenges stemming from the inability to exactly 
sample surrogate trajectories $\targetTraj \sim \emDist$. On the other hand, 
\citet{BurknerSurrogate,garegnani2021NoisyMCMC} appear to implicitly assume the use of 
finite-dimensional surrogate models for which sampling trajectories is straightforward.
See \Cref{sec:computation} for additional computational details.

In the modular Bayes literature, there is a wide body of work 
on the cut posterior, which is equivalent to the EP 
when a fully Bayesian reference model is considered
\citep{PlummerCut,cutInference,moduleModels}.
Various papers have justified the cut posterior as a reverse KL divergence minimizer, 
akin to \Cref{prop:kl-cut-op} \citep{cutVar,cutVar2,moduleModels}. 

The hierarchical sampling view of the EP in \Cref{eq:ep-prob-model} also 
corresponds to a Bayesian multiple imputation algorithm, typically applied
in missing data problems \citep{multipleImputationMedical,missingData}.
The notion of aggregating multiple posterior distributions is also used in 
contexts other than modular inference, including for robustness to model 
misspecification \citep{BayesBag,BayesBag2}.

\section{Approximating the Expected Posterior} \label{sec:approximating-ep}
Having justified the EP as the baseline target distribution for modular uncertainty 
propagation, we now turn to the practical question of conducting EP-based
inference. While exact inference is typically infeasible, two 
natural strategies are to
(1) target an alternative distribution that approximates the EP but is more amenable
to standard inference algorithms, and (2) employ an approximate 
inference algorithm directly targeting the EP (see \Cref{sec:computation} for the latter). 
Previous studies have adopted the former approach, eschewing the EP in favor 
of approximations of the \textit{unnormalized} density surrogate $\postEm$
\citep{StuartTeck1,StuartTeck2,VehtariParallelGP}.
However, these works stop short of studying how this approximation
relates to the EP. In this section, we provide multiple characterizations of this approximation,
termed the  \textit{expected unnormalized posterior (EUP)}, and provide analytical results 
illustrating how the EP and EUP can differ.

\subsection{The Expected Unnormalized Posterior} \label{sec:eup}
Unlike $\postNormEm(\Par)$, the unnormalized density surrogate
$\postEm(\Par)$ depends only on the single-point prediction $\targetEm(\Par)$
rather than the full emulator $\targetEm$. The EUP is defined by computing
a pointwise expectation of $\postEm(\Par)$ and then normalizing post-hoc:
\begin{equation}
\postApproxEUPNorm(\Par) \Def 
\frac{\emE \left[\postEm(\Par) \right]}{\int_{\parSpace} \emE\left[\postEm(\Par) \right] \d\Par}
= \frac{\emE\left[\postEm(\Par) \right]}{\emE[\normCstEm]}. \label{eq:post-approx-EUP} 
\end{equation}
The equality in \Cref{eq:post-approx-EUP} follows from
changing the order of integration, courtesy of Tonelli's theorem \citep{StuartTeck1}. The EUP
is a marginal of the joint distribution 
$\postDens(\Par; \targetTraj) / \emE[\normCst(\targetEm)]$. 
As compared to the analogous EP joint
$\postDens(\Par; \targetTraj) / \normCst(\targetTraj)$,
we see that the EUP replaces the normalizing function $\normCst(\targetTraj)$ with
the global point estimate $\emE[\normCst(\targetEm)]$. The $\targetTraj$-marginal of
this joint is proportional to $\emDist(\d\targetTraj) \normCst(\targetTraj)$, implying the
joint is not an element of $\qSpaceCut$, the space of cut distributions defined in \Cref{eq:cut-dists}.
In other words, the joint probability model implied by the EUP does not preserve $\emDist$ as the 
marginal for $\targetTraj$, instead allowing the observational data $\obs$ to alter this marginal.

Like the EP, the EUP admits
both posterior mixture and marginal likelihood interpretations. Starting with the latter, 
the EUP arises as the marginal posterior $p(\Par \given \obs)$ under the hierarchical model
\begin{equation}
\targetTraj  \sim \emDist, \qquad 
\Par \sim \priorDens, \qquad
\obs \given \targetTraj, \Par \sim \lik(\Par; \targetTraj, \d\obs).
\label{eq:eup-prob-model}
\end{equation}
This yields 
\begin{align*}
&\postApproxEUPNorm(\Par) \propto \priorDens(\Par) \likApproxEUP(\Par; \obs),
&&\likApproxEUP(\Par; \obs) \Def \int \lik(\Par; \targetTraj, \obs) \emDist(\d\targetTraj), 
\end{align*}
providing an analog to the EP characterization in \Cref{eq:lik-approx-EP}. In contrast
with $\likApproxEP(\Par; \obs)$, the EUP marginal likelihood does not include 
$\normCst(\targetTraj)$ in the integrand. Analogous to \Cref{eq:ep-approx}, the EUP can also be 
written as a posterior mixture
\begin{equation}
\postApproxEUPNorm(\Par) = \int \postDensNorm(\Par; \targetTraj) \emDist(\d\targetTraj \given \obs),
\label{eq:eup-mixture} 
\end{equation}
where $\emDist(\d\targetTraj \given \obs) \propto \normCst(\targetTraj) \emDist(\d\targetTraj)$.
Hence, the plug-in mean, EP, and EUP can all be written in this mixture form with the respective
weights $\delta_{\Em{\bar{\targetTraj}}}(\d\targetTraj)$, $\emDist(\d\targetTraj)$, and $\emDist(\d\targetTraj \given \obs)$.

\subsection{Comparison of the EP and EUP}
Notice in \Cref{eq:post-approx-EUP} that the EUP is a ratio estimator, and differs from 
the EP due to the nonlinearity of the normalization operation. The EUP can thus 
be derived from the EP by invoking two approximations: 
\emph{(i)} treating $\postEm(\Par)$ and $\normCstEm^{-1}$ as independent; and 
\emph{(ii)} assuming $\emE[\normCstEm^{-1}] \approx \emE[\normCstEm]^{-1}$.
The following result quantifies the effect of these two approximations.

\begin{prop} \label{prop:ep-eup-pw-err}
The pointwise difference between the EP and EUP is given by
\begin{equation}
\postApproxEP(\Par) - \postApproxEUPNorm(\Par)
= \emE[\postEm(\Par)] \jgap + \Cov[\postEm(\Par), \normCstEm^{-1}],
\label{eq:ep-eup-pw-err}
\end{equation}
where $\jgap \Def \emE[\normCstEm^{-1}] - \emE[\normCstEm]^{-1}$
is the ``Jensen gap.''
\end{prop}

By Jensen's inequality $\jgap \geq 0$, implying that the Jensen gap represents a 
$\Par$-independent positive bias in the difference between EP and EUP, modulated by
$\emE[\postEm(\Par)]$. Since both distributions integrate to one, any positive 
biases must be balanced by negative biases at other values of $\Par$.
The second term in \Cref{eq:ep-eup-pw-err} will be negative for ``influential'' $\Par$ 
values---those with larger realizations of $\postEm(\Par)$ highly correlated with larger values of 
$\normCstEm$. The influence of a parameter value will typically increase when 
$\postEm(\Par)$ is large on average, highly variable, and is positively correlated with other 
$\postEm(\Par^\prime)$. The latter property may be satisfied, for example, when using 
a GP emulator with a long lengthscale. Based on this logic, we expect the EUP to inflate 
influential regions and depress non-influential regions, relative to the EP. As demonstrated in
experiments (\Cref{sec:vsem}), this can manifest as the EUP being highly peaked in regions with 
significant surrogate uncertainty. The pointwise error in 
\Cref{prop:ep-eup-pw-err} can be integrated to obtain the following $L_1$ bound 
between $\postApproxEP$ and $\postApproxEUPNorm$.

\begin{prop} \label{prop:ep-eup-TV-err}
Let $\jgap$ be defined as in \Cref{prop:ep-eup-pw-err}. Then,
\begin{equation}
\norm{\postApproxEP - \postApproxEUPNorm}_{L_1(\parSpace)}
\leq \emE[\normCstEm] \jgap + \int \abs{\Cov[\postEm(\Par), \normCstEm^{-1}]} d\Par
\label{eq:ep-eup-TV-err}
\end{equation}
\end{prop}

Decreasing the variance of $\normCstEm$ will shrink both terms in 
\Cref{eq:ep-eup-TV-err}. This may occur when there is little 
uncertainty in $\targetEm$, the unnormalized posterior $\postDens(\cdot; \targetTraj)$
is insensitive to $\targetTraj$, or $\normCstEm$ is insensitive to the variability 
in $\postEm$. In the first case, note that if $\targetEm$ is heavily 
concentrated around its mean, then both the EP and EUP will closely resemble
the plug-in mean approximation in \Cref{eq:mean-approx}.
In special cases, the two terms in \Cref{eq:ep-eup-TV-err} perfectly balance so that 
the EP and EUP agree. For example, this occurs when 
$\postEm(\Par) = \omega g(\Par)$, where $\omega$ is a random constant and 
$g(\Par)$ a deterministic function. 

\subsubsection{Conceptual Considerations} \label{sec:conceptual}
It should be emphasized that the correct choice of uncertainty propagation method  
may be problem-dependent. We view the variational justification for the EP provided by
\Cref{prop:EP-variational} as a general guide, but particular conceptual considerations
may take precedence in certain situations. The EUP is a natural choice when $\targetEm$
is viewed as a latent variable that forms part of the data-generating process for the observational
data $\obs$, leading to the hierarchical model in \Cref{eq:eup-prob-model}. Our focus is 
on situations where the randomness in $\targetEm$ is primarily epistemic---in principle it could be 
reduced via further evaluations of the simulator. The EP effectively follows from the view that this 
epistemic uncertainty is external to the data-generating process. Another important consideration
is that the EP is a true cut posterior distribution, while the EUP incorporates the data $\obs$ into
the uncertainty propagation method. In this sense, the EUP is ``partially modular''---the surrogate 
is fit using only simulator data, but the observational data alters the weights assigned to surrogate
trajectories in the second stage.

\subsection{EUP Examples}
We return to \Cref{ex:fwd-em,ex:ldens-em} to highlight concrete characterizations of the EUP
in two common practical settings. 

\begin{ex}[Forward Model GP Surrogate]
\label{ex:fwd-em-cont}
Consider the setup from example \Cref{ex:fwd-em} with approximate density 
$\postEm(\Par) = \priorDens(\Par)\Gaussian(\obs \mid \targetEm(\Par), \likPar)$. We now 
additionally assume that the surrogate has Gaussian pointwise predictions 
$\targetEm(\Par) \sim \GP(\emMean(\Par), \emVar(\Par))$. Such predictions might come from 
a (potentially multi-output) GP emulator, though note that the EUP is defined only by the 
pointwise predictions, and ignores any correlational structure in the surrogate.
This setup is commonly considered in the Bayesian 
inverse problem literature \citep{Surer2023sequential,weightedIVAR,StuartTeck2,GP_PDE_priors,CES,
idealizedGCM,villani2024posteriorsamplingadaptivegaussian,hydrologicalModel,hydrologicalModel2}.

Under these assumptions, the EUP assumes the form
\begin{equation}
\postApproxEUPNorm(\Par) \propto \priorDens(\Par) \Gaussian(\obs \mid \emMean(\Par), \likPar + \emVar(\Par)),
\label{eq:EUP-Gaussian-fwd}
\end{equation}
following from the formula for the convolution of two Gaussians \citep{StuartTeck2}.
In this context, the uncertainty propagation admits a data space interpretation, with
$\gpKer(\Par)$ inflating the observation covariance. This implies reversion to the prior density 
$\postApproxEUPNorm(\Par) \to \priorDens(\Par)$ as $\emVar(\Par) \to \infty$. Note that it is not correct to 
think of \Cref{eq:EUP-Gaussian-fwd}  as simply inflating the plug-in mean approximation in regions with 
high uncertainty. Since the Gaussian likelihood is bounded above, when $\emMean(\Par)$ and $\obs$ are close, 
then an increase in $\emVar(\Par)$ may actually deflate the density at $\Par$.
\end{ex}

\begin{ex}[Log-Density GP Surrogate]
\label{ex:ldens-em-cont}
Consider the setup from \Cref{ex:ldens-em} where $\postEm(\Par) = \Exp{\targetEm(\Par)}$.
We again assume an emulator with Gaussian predictions $\targetEm(\Par) \sim \GP(\emMean(\Par), \emVar(\Par))$. 
This encompasses both the log-likelihood and log-posterior emulation settings; in the former case, 
$\emMean(\Par)$ is the sum of $\log \priorDens(\Par)$ and the mean of the log-likelihood emulator.
This setup has been considered in several applications
\citep{VehtariParallelGP,FATES_CES,trainDynamics,quantileApprox,
ActiveLearningMCMC,FerEmulation,StuartTeck1,StuartTeck2,random_fwd_models,
GP_PDE_priors,OakleyllikEm,JosephMinEnergy,AlawiehIterativeGP,gpEmMCMC}.

Under these assumptions, the pushforward predictive distribution for $\postEm(\Par)$ is log-normal,
$\postEm(\Par) \sim \LN(\gpMean(\Par), \emVar(\Par))$, implying
\begin{equation}
\postApproxEUPNorm(\Par) 
\propto \Exp{\gpMean(\Par) + \frac{1}{2}\emVar(\Par)}
= \postApproxMean(\Par) \Exp{\frac{1}{2}\emVar(\Par)}.
\label{eq:EUP-Gaussian-ldens}
\end{equation} 
As opposed to the preceding example, the EUP in this context is correctly interpreted as 
pointwise inflation of the plug-in mean approximation in accordance with the surrogate uncertainty. 
At two points with equal predictive mean $\gpMean$, the EUP always assigns higher density to the 
more uncertain location. Also in contrast to the previous example, \Cref{eq:EUP-Gaussian-ldens}
does not exhibit prior reversion as $\emVar(\Par)$ increases at a particular point; as 
$\postApproxEUPNorm(\Par)  \to \infty$ as $\emVar(\Par) \to \infty$.
Notably the magnitude of the uncertainty inflation scales with the exponentiated predictive 
variance, making the EUP susceptible to extreme concentration in small regions exhibiting high uncertainty.
For example, if $\emSD(\Par) = 2$,
$\emMean(\Par^\prime) = \emMean(\Par)$, and 
$\emSD(\Par^\prime) = 2 \emSD(\Par)$, then 
$\{\postApproxEUPNorm(\Par^\prime)/\postApproxEUPNorm(\Par)\} / 
\{\postApproxNormMean(\Par^\prime)/\postApproxNormMean(\Par)\} \approx 400$. 
The $2\times$ difference in surrogate standard deviation translates to a $400\times$ difference in 
density approximation, relative to the plug-in mean. This undesirable behavior, also noted by 
\citet{VehtariParallelGP}, can be viewed partially as a consequence of the dominating effect
of the second term in \Cref{eq:ep-eup-pw-err} at points where the distribution of 
$\postEm(\Par)$ is heavy-tailed.
As illustrated in the experiment in \Cref{sec:vsem}, the EP tends to be more robust, 
though not immune, to the sensitivity exhibited by posterior estimates under GP log-density surrogates.
\end{ex}

\paragraph{Related work.} The EUP is proposed in the forward model emulator setting 
in \citet{BilionisBayesSurrogates}, motivated by the extended parameter space viewpoint
in \Cref{eq:eup-prob-model}. \citet{StuartTeck2,CES} also note this perspective
in the particular Gaussian setting of \Cref{eq:EUP-Gaussian-fwd}.
In \citet{SinsbeckNowak}, the EUP is justified as the distribution 
$q$ that minimizes $\emE\left[\norm{\postEm - q}_{L_2(\parSpace)}^2 \right]$.
\citet{StuartTeck1,StuartTeck2,VehtariParallelGP} also highlight this Bayesian 
decision theoretic justification. 
In contrast with the EP, the optimality is only guaranteed for the estimate of the 
\textit{unnormalized} posterior.
As shown in \Cref{prop:EP-variational}, the EP is the minimizer when the loss
is defined with respect to the normalized distributions.
The EUP is referred to as the ``marginal'' approximation in 
\citet{StuartTeck1,StuartTeck2,random_fwd_models,TeckHyperpar},
which establish error bounds with respect to the true posterior in Hellinger distance.
\citet{VehtariParallelGP} highlight pathological 
behavior of the EUP for GP log-density emulators, and recommend 
against its use in this setting. To our knowledge, \citet{BurknerSurrogate}
is the only previous work to directly compare the EP and EUP (which 
they call the ``expected likelihood''). Their comparison is limited to 
numerical results in case studies involving forward model emulators.
The closed-form EUP expressions in 
\Cref{eq:EUP-Gaussian-ldens,eq:EUP-Gaussian-fwd} have been 
noted in a variety of studies 
\citep{StuartTeck1,StuartTeck2,VehtariParallelGP,weightedIVAR,
GP_PDE_priors,Surer2023sequential,Takhtaganov2018AdaptiveBayesianGP}. 
The EUP has seen 
use in various applications involving both forward model 
and log-density emulators 
\citep{weightedIVAR,GP_PDE_priors,CES,idealizedGCM,
villani2024posteriorsamplingadaptivegaussian,hydrologicalModel,hydrologicalModel2}.
 
\section{Approximate Computation for the Expected Posterior} \label{sec:computation}
The previous section explored the EUP as an approximation to the EP, demonstrating when 
the two distributions may deviate. We now seek a more direct route to EP-based
inference, and introduce an approximate MCMC algorithm towards this end. 
We start by clarifying the difficulties associated with EP computation.

\subsection{Sampling Trajectories}
In light of the hierarchical model in \Cref{eq:ep-prob-model}, the following 
algorithm can in principle be applied to directly sample $\postApproxEP$.  

\begin{algorithm}[H]
    \caption{Direct sampling from \texorpdfstring{$\postApproxEP$}{EP}}
    \label{alg:ep}
    \begin{algorithmic}[1]
    \Function{sampleEP}{$\postNormEm, \NSample, M$}     
        \For{$\sampleIndex \gets 1, \dots, \NSample$} 
        		\State $\targetTraj^{(\sampleIndex)} \sim \emDist$ \Comment{Sample emulator trajectory}
		\State $\Par^{(\sampleIndex, 1)}, \dots, \Par^{(\sampleIndex, M)} \sim \postDensNorm(\cdot; \targetTraj^{(\sampleIndex)})$ \Comment{Sample posterior given trajectory}
	\EndFor
	\State \Return $\{\Par^{(\sampleIndex, m)}\}_{1 \leq \sampleIndex \leq \NSample, \ 1 \leq m \leq M}$
	\EndFunction
    \end{algorithmic}
\end{algorithm}
If one sample is drawn from each posterior trajectory (i.e., $M=1$) then the resulting samples are 
independent. Otherwise, \Cref{alg:ep} produces dependent samples identically distributed according to
$\postApproxEP$. In practice, directly sampling $\Par \given \targetTraj \sim \postDensNorm(\cdot; \targetTraj)$
is rarely possible, so this inner sampling step is replaced by an MCMC algorithm.
The resulting sampling scheme is sometimes called \textit{Metropolis within Monte Carlo (MwMC)} 
\citep{garegnani2021NoisyMCMC}. MwMC has the downside of requiring $\NSample \gg 1$
MCMC runs, but this may be less of an issue in modern parallel computing environments \citep{BurknerSurrogate}. 
Moreover, methods have been developed to reduce the number of sampled trajectories $\NSample$ 
required to adequately characterize the EP \citep{BurknerTwoStep}. 

Another issue is the outer sampling step, which requires simulating surrogate 
trajectories $\targetEm \sim \emDist$. While not a problem for finite-dimensional 
surrogate models (e.g., linear models), this presents major challenges for surrogates 
derived from GPs. Given the popularity of GP surrogates, 
this computational bottleneck must be resolved for the EP to be broadly accessible in 
surrogate-based Bayesian workflows. Standard remedies suffer from poor scalability,
limited applicability, or bespoke numerical implementations. For example, naive 
approximations that discretize $\parSpace$ are limited to low-dimensional settings.
Finite-rank GP approximations offer an alternative, but are dependent on the 
particular form of the surrogate \citep{pathwiseConditioning}. In theory, one 
could retain an infinite-dimensional GP representation by constructing 
GP trajectories ``just-in-time'' within the MCMC algorithm; that is, iteratively
condition the GP at each value of $\Par$ visited within the MCMC run.
However, this approach is well-known to suffer from significant numerical 
instability \citep{pathwiseConditioning}. Finally, we highlight the method of 
\citet{trainDynamics}, which is to our knowledge the only existing work to 
attempt EP-based inference with a GP surrogate. 
Their method consists of approximating GP 
trajectories by sampling the surrogate at a finite grid of points, 
and then approximating the trajectory as the GP mean, conditional on the 
sampled values at these points. This approach has the downside of 
requiring the difficult choice of an appropriate conditioning set.

\subsection{Exact Metropolis-within-Gibbs} \label{sec:mwg}
Given the downsides of the MwMC approach, a natural alternative is to seek an MCMC
algorithm that targets the EP asymptotically. This offers the potential for computational 
savings---running one chain instead of a large ensemble---and avoids the need
to sample trajectories. We first introduce an exact, but impractical, Metropolis-within-Gibbs (MwG) 
algorithm that targets the EP. We then consider several approximations that yield practical approximate
samplers.

We aim to target the joint distribution 
$\emDist(d\targetTraj) \postDens(\Par; \targetTraj)/\normCst(\targetTraj) d\Par$, which admits $\postApproxEP$ 
as a marginal. We consider MwG schemes that alternate between $\Par$ and $\targetTraj$ updates, which 
must leave their respective conditional distributions invariant. The $\Par$ update must leave 
$\postDens(\Par; \targetTraj) d\Par$ invariant, which is accomplished with a standard Metropolis-Hastings (MH)
step. The $\targetTraj$ update must leave 
 $\emDist(d\targetTraj) \postDens(\Par; \targetTraj)/\normCst(\targetTraj)$ invariant, requiring more 
 care due to the fact that $\targetTraj$ may be infinite-dimensional.
 Proceeding in the spirit of function space MCMC methods \citep{functionSpaceMCMC}, we consider an MH step with 
a $\emDist$-reversible proposal distribution $\propDistTarget(\targetTraj, \cdot)$, resulting in 
the acceptance probability 
\begin{equation}
\accProbMH_{\target}(\targetTraj, \tilde{\targetTraj})
= \min\left(1, \frac{\postDens(\Par; \tilde{\targetTraj})}{\postDens(\Par; \targetTraj)} \cdot 
\frac{\normCst(\targetTraj)}{\normCst(\tilde{\targetTraj})} \right)
= \min\left(1, \frac{\postDensNorm(\Par; \tilde{\targetTraj})}{\postDensNorm(\Par; \targetTraj)} \right).
\label{eq:MH-ratio-mwg}
\end{equation}
Henceforth, we consider the particular case where $\emDist$ is a Gaussian measure $\Gaussian(\emMean, \gpKer)$. 
\footnote{More generally, the emulator predictive distribution itself need not be Gaussian. By simply re-defining
$\target$, $\emDist$ can be the pushforward of an underlying Gaussian measure.} This covers many 
practical applications where the surrogate is constructed as some function of a GP, a setting that currently 
raises challenges for EP-based inference. Generalization to the non-Gaussian setting is straightforward when 
$\emDist$ is finite-dimensional. In the Gaussian setting, a well-known $\emDist$-reversible proposal is given
by the preconditioned Crank-Nicolson (pCN; \citet{functionSpaceMCMC}) update 
\begin{equation}
\tilde{\targetTraj} \Def \gpMean + \pcnCor (\targetTraj  - \gpMean) + \sqrt{1 - \pcnCor^2} \xi, 
\qquad \xi \sim \Gaussian(0, \gpKer). \label{eq:pcn-proposal}
\end{equation}
\Cref{alg:mwg} summarizes the complete MwG algorithm with the pCN proposal. 
 
\begin{algorithm}[H]
    \caption{Metropolis-within-Gibbs for EP (single iteration)}
    \label{alg:mwg}
    \begin{algorithmic}[1]
    \State \textbf{Input:} Current state $(\Par, \targetTraj)$
    \State \textbf{Output:} Updated state $(\nextstate{\Par}, \nextstate{\targetTraj})$
    \State $\xi \sim \Gaussian(0, \gpKer), v_{\targetTraj} \sim \mathrm{Unif}(0, 1)$
    \State $\tilde{\targetTraj} \gets \gpMean + \pcnCor (\targetTraj  - \gpMean) + \sqrt{1 - \pcnCor^2} \xi$ \Comment{pCN proposal}
    \State $\alpha_{\targetTraj} \gets \min\left\{1, \postDensNorm(\Par; \tilde{\targetTraj}) / \postDensNorm(\Par; \targetTraj) \right\}$ 
    \State $\nextstate{\targetTraj} \gets \tilde{\targetTraj} \indicator{v_{\targetTraj} \leq \alpha_{\targetTraj}} + \targetTraj \indicator{v_{\targetTraj} > \alpha_{\targetTraj}}$ \Comment{$\targetTraj$ update}
    \State $\propPar \sim \propDistPar(\Par, \cdot), v_\Par \sim \mathrm{Unif}(0,1)$
    \State $\alpha_{\Par} \gets \min\left\{1, \postDens(\propPar; \nextstate{\targetTraj}) / \postDens(\Par; \nextstate{\targetTraj}) \right\}$
    \State $\nextstate{\Par} \gets \propPar \indicator{v_\Par \leq \alpha_{\Par}} + \Par \indicator{v_\Par > \alpha_{\Par}}$ \Comment{$\Par$ update}
    \end{algorithmic}
\end{algorithm}

This algorithm generally cannot be implemented due to the intractable normalizing 
constant ratio $\normCst(\targetTraj)/\normCst(\tilde{\targetTraj})$ in \Cref{eq:MH-ratio-mwg}. 
At first glance, it appears that techniques from the doubly intractable MCMC literature 
\citet{doublyIntractableReview,exchangeAlg} may circumvent this issue, but such methods
 typically require the ability to directly sample $\postDensNorm(\cdot; \tilde{\targetTraj})$,
which is infeasible in this context. An exact pseudo-marginal implementation is possible, but requires
an unbiased estimator for $\normCst(\targetTraj)^{-1}$. While this is in principle possible using truncated
random series (e.g., \citet{yvesCut}), such methods are difficult to implement in practice. We instead
seek practical approximate heuristics that exploit the structure of the pCN proposal. 

\subsection{Random Kernel Metropolis-Hastings}
We now consider approximating the $\targetTraj$ update in the MwG scheme, leaving
the $\Par$ update unchanged. Intuitively, by setting the pCN correlation parameter $\pcnCor \approx 1$
the proposal $\tilde{\targetTraj}$ will be close to the current state $\targetTraj$, increasing the acceptance
probability $\accProbMH_{\target}(\targetTraj, \tilde{\targetTraj})$. Following this intuition, we consider invoking the 
simple approximation $\postDensNorm(\Par; \tilde{\targetTraj}) / \postDensNorm(\Par; \targetTraj) \approx 1$, 
thus removing the accept-reject correction and instead allowing the $\targetTraj$ chain 
to follow a random walk. The notion of slowing the mixing speed of the $\targetTraj$-chain
is noted in \citet{PlummerCut}, though it is discounted on the basis of inducing artificially slow mixing.
By contrast, this slow mixing is quite natural in the present high-dimensional context, where large values of 
$\pcnCor$ are already typically required to prevent excessively high rejection rates 
\citep{functionSpaceMCMC, corrPM}. The resulting algorithm, which we call 
\textit{random kernel preconditioned Crank-Nicolson (RKpCN)}, is summarized in \Cref{alg:rk-pcn-finite}.
Though it is stated for conceptual clarity in infinite dimensions, we emphasize that the algorithm can be 
implemented exactly by only realizing finite-dimensional projections of the functions. 
See \Cref{alg:rk-pcn-infinite} in \Cref{app:mcmc} for details. We illustrate the effect of the RKpCN acceptance 
probability approximation through experiments; a complete theoretical analysis is beyond the scope of this paper.

\begin{rmk}
The RKpCN scheme is derived by invoking the approximation
$\postDensNorm(\Par; \tilde{\targetTraj}) / \postDensNorm(\Par; \targetTraj) \approx 1$.
A natural alternative is to consider the approximation 
$\normCst(\targetTraj) / \normCst(\tilde{\targetTraj}) \approx 1$,
which yields the acceptance probability
\begin{equation}
\accProbMH_{\target}(\targetTraj, \tilde{\targetTraj})
= \min\left(1, \frac{\postDens(\Par; \tilde{\targetTraj})}{\postDens(\Par; \targetTraj)}\right).
\label{eq:MH-ratio-cpm}
\end{equation}
This is precisely a correlated psuedo-marginal algorithm that exactly targets $\postApproxEUPNorm$.
This provides an alternative algorithmic derivation of the EUP based on a heuristic modification of 
an algorithm targeting the EP.
See \citet{pseudoMarginalMCMC,pseudoMarginalEfficiency,corrPM}
for background on pseudo-marginal MCMC.
\end{rmk}

\begin{algorithm}[H]
    \caption{Random Kernel pCN (single iteration)}
    \label{alg:rk-pcn-finite}
    \begin{algorithmic}[1]
    \State \textbf{Input:} Current state $(\Par, \targetTraj)$
    \State \textbf{Output:} Updated state $(\nextstate{\Par}, \nextstate{\targetTraj})$
    \State $\xi \sim \Gaussian(0, \gpKer)$
    \State $\nextstate{\targetTraj} \gets \gpMean + \pcnCor (\targetTraj  - \gpMean) + \sqrt{1 - \pcnCor^2} \xi$ \Comment{$\targetTraj$ update}
    \State $\propPar \sim \propDistPar(\Par, \cdot), v_\Par \sim \mathrm{Unif}(0,1)$
    \State $\alpha_{\Par} \gets \min\left\{1, \postDens(\propPar; \nextstate{\targetTraj}) / \postDens(\Par; \nextstate{\targetTraj}) \right\}$
    \State $\nextstate{\Par} \gets \propPar \indicator{v_\Par \leq \alpha_{\Par}} + \Par \indicator{v_\Par > \alpha_{\Par}}$ \Comment{$\Par$ update}
    \end{algorithmic}
\end{algorithm}
 
\paragraph{Related work.}
Both \citep{garegnani2021NoisyMCMC} and \citet{BurknerSurrogate} propose MwMC schemes
for EP-based inference. They appear to implicitly assume the use of finite-dimensional emulators,
as difficulties related to sampling trajectories are not addressed.
\citet{FerEmulation} utilize a noisy MH algorithm in which $\nextstate{\targetTraj}_{\Par}$ and 
$\nextstate{\targetTraj}_{\tilde{\Par}}$ are independently sampled at each step.

There has been interest in the modular Bayes community in designing approximate MCMC 
schemes as an alternative to MwMC for cut posterior inference. \citet{PlummerCut} describes 
the implementation of such an algorithm in the WinBUGS software, referred to as the 
\textit{naive cut algorithm}. The author shows that this algorithm does not admit the cut posterior 
as an invariant distribution, and moreover that the implicit target distribution depends on the 
particular Markov kernels chosen to perform the updates.
The paper proposes a solution to 
improve the approximation using tempered transitions. Subsequent work has considered more 
sophisticated algorithms that seek to explicitly estimate the intractable normalizing constant ratios
\citep{SAACut} or utilize coupling approaches based on unbiased telescoping sum estimators \citep{yvesCut}.
Our proposed method can be viewed as a version of the cut algorithm operating in function space.
We opt to avoid normalizing constant estimation and instead slow down the mixing of the 
$\targetTraj$-chain to control the approximation error.

\section{Numerical Experiments} \label{sec:experiments}
We consider several numerical experiments, with the dual aims of \emph{(i)} comparing the EP, EUP, and
plug-in mean approximation under various surrogate modeling setups, and \emph{(ii)} assessing the 
quality of the RKpCN approximation to the EP. For the latter, we evaluate the
RKpCN algorithm at different values of the correlation parameter $\pcnCor$. 
We compare these MCMC approximations against the EUP as well as the RKpCN scheme with
$\pcnCor = 0$. The latter implies sampling independent realizations of the surrogate at 
each MCMC iteration, similar to the approach in \citet[Section 3]{PlummerCut}. Although this $\pcnCor = 0$
case is technically a RKpCN algorithm, we refer to it in experiments as the \textit{independent cut} scheme,
given that it lacks the defining quality (large $\pcnCor$) that justifies the approximation used in RKpCN.

\subsection{Linear Gaussian Example} \label{sec:linear-gaussian}
We start with a toy linear Gaussian model in which the exact posterior, EP, and EUP are all 
Gaussian and available in closed-form. A similar example is considered in
\citet{garegnani2021NoisyMCMC}. Consider the linear Gaussian inverse problem
\begin{align}
\obs &= \fwdLin \Par + \noise,
&&\noise \sim \Gaussian(0, \noiseCov) \label{eq:lin-Gauss-inv-prob} \\
\Par &\sim \Gaussian(\priorMean, \priorCov). \nonumber
\end{align}
The exact posterior is $\Par \given \obs \sim \Gaussian(\postMean, \postCov)$, where
\begin{align}
\postMean &= \postCov \left(\fwdLin^\top \noiseCov^{-1}\obs + \priorCov^{-1}\priorMean \right) \label{eq:lin-Gauss-post-moments} \\
\postCov &= \left(\fwdLin^\top \noiseCov^{-1}\fwdLin + \priorCov^{-1} \right)^{-1}. \nonumber
\end{align}
We consider a forward model emulator of the form
\begin{align}
\targetEm(\Par) \sim \Gaussian(\fwdLin\Par + \emBias, \emCov) \label{eq:lin-Gauss-em}, 
\end{align}
corresponding to an additive Gaussian shift of the true model, with bias $\emBias$.
The covariance $\emCov$ quantifies the surrogate uncertainty regarding the magnitude of the bias.
In this toy example, the surrogate randomness does not vary with $\Par$, in contrast to the 
emulators considered in our subsequent, more realistic, experiments.
The following result characterizes the EP and EUP in this setting.

\begin{prop}
Under the linear Gaussian model in \Cref{eq:lin-Gauss-inv-prob} with the emulator in \Cref{eq:lin-Gauss-em},
the EP is given by $\postApproxEP(\Par) = \Gaussian(\Par \given \postMeanEP, \postCovEP)$, where
for $\epGain \Def \postCov \fwdLin^\top \noiseCov^{-1}$,
\begin{align}
\postMeanEP &= \postMean - \epGain \emBias,
&&\postCovEP = \postCov + \epGain \emCov \epGain^\top.
\label{eq:lin-gauss-ep}
\end{align}
The quantities $\postMean, \postCov$ are the 
exact posterior moments given in \Cref{eq:lin-Gauss-post-moments}. Under the same setup, the 
EUP is given by $\postApproxEUP(\Par) = \Gaussian(\Par \given \postMeanEUP, \postCovEUP)$, where
\begin{align}
&\postMeanEUP = \postCovEUP \left(\fwdLin^\top \eupNoiseCov^{-1}\eupObs + \priorCov^{-1}\priorMean \right),
&&\postCovEUP = \left(\fwdLin^\top \eupNoiseCov^{-1}\fwdLin + \priorCov^{-1} \right)^{-1}
\label{eq:lin-gauss-eup}
\end{align}
and $\eupNoiseCov \Def \noiseCov + \emCov$, $\eupObs \Def \obs - \emBias$.
\label{eq:lin-gauss-posts}
\end{prop}

\subsubsection{Analytical Analysis}
Let $G = USV^\top$ denote the singular value decomposition of $G$, with the singular values $(s_j)$ sorted
in descending order. The right singular vectors $v_j$ associated with large $s_j$ correspond to directions in 
parameter space that are well-informed by the data. We investigate the posterior mean and covariance of 
the EUP and EP in the $V$ basis to understand the influence of the surrogate noise. To simplify matters, consider
the special case $\noiseCov = \sigma^2 I$, $\priorCov = c_0^2 I$, $\emCov = q^2 I$. The following results give
the expressions for the exact, EUP, and EP posterior moments in the $V$ basis.
\begin{prop} \label{prop:linear-gaussian-analytical}
Under the above assumptions $\postCov$, $\postCovEUP$, and $\postCovEP$ are diagonalized in the 
$V$ basis, with respective eigenvalues given by 
\begin{align}
\lambda_j = \frac{c_0^2}{1 + \frac{c_0^2 s_j^2}{\sigma^2}}, \qquad
\lambda_j^{\eup} = \frac{c_0^2}{1 + \frac{c_0^2 s_j^2}{\sigma^2 + q^2}}, \qquad
\lambda_j^{\ep} = \lambda_j + \frac{q^2 s_j^2 c_0^4}{\sigma^4}.
\end{align}
Moreover, the posterior means of each distribution can be written as linear combination of the $v_j$. The 
coefficients in these linear combinations in directions with $s_j \neq 0$ are given by
\begin{align*}
&\alpha_j = \frac{\lambda_j s_j}{\sigma_j^2} \langle y, u_j \rangle + \frac{\lambda_j}{c_0^2} \langle m_0, v_j \rangle \\ 
&\alpha_j^{\eup} = \frac{\lambda_j s_j}{\sigma^2 + q^2} \langle y - r, u_j \rangle + \frac{\lambda_j}{c_0^2} \langle m_0, v_j \rangle,
&&\alpha_j^{\ep} = \alpha_j - \frac{c_0^2 s_j}{\sigma^2} \langle r, u_j \rangle.
\end{align*}
\end{prop}
The behavior of the EUP and EP as the surrogate uncertainty increases is markedly different. As $q \to \infty$, the EUP 
reverts to the $\Gaussian(\priorMean, \priorCov)$ prior, while the EP tends toward a flat distribution centered
on $\postMeanEP$. In the non-asymptotic regime, the surrogate uncertainty inflates both the EP and EUP 
variance most significantly along directions well-informed by the data ($v_j$ with $s_j$ large). These 
trends are illustrated for a concrete numerical example in \Cref{fig:lin-gauss-svd-plots}. Note that $\postMeanEP$
is independent of $\emCov$ and thus does not vary with the level of surrogate uncertainty.

\subsubsection{A Deconvolution Problem} \label{sec:deconvolution}
We now consider a concrete instantiation of the linear Gaussian inverse problem, representing
deconvolution for a one-dimensional signal. The parameter $\Par \in \R^{100}$ of interest is 
a discretized signal over the domain $\parSpace = [0,100]$. The forward operator $G$ is the composition
of a linear convolution with a Gaussian kernel, with an observation operator that selects every 
fourth grid point. The data space is thus of dimension $\dimObs=25$. We consider $\Sigma = \sigma^2 I$
with $\sigma = 0.2$ and define $\priorCov$ as a Gaussian kernel matrix, encoding a smoothness 
assumption for the signal.

\begin{figure}[htbp]
    \centering

    \begin{subfigure}[b]{\textwidth}
        \centering
        \begin{subfigure}[b]{0.45\textwidth}
            \includegraphics[width=\textwidth]{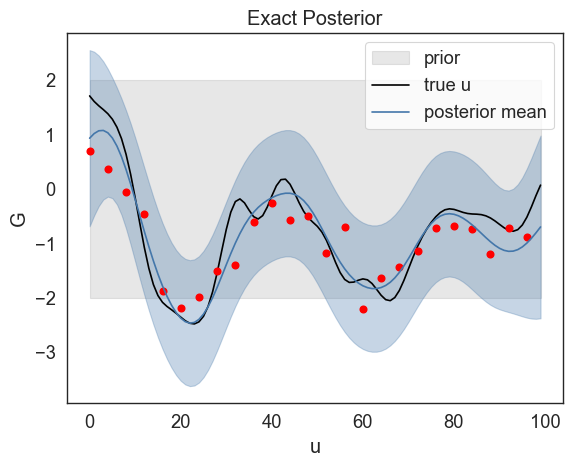}
             \caption{Exact Posterior}
             \label{fig:lin-gauss-exact-post}
        \end{subfigure}
        \hfill
        \begin{subfigure}[b]{0.45\textwidth}
            \includegraphics[width=\textwidth]{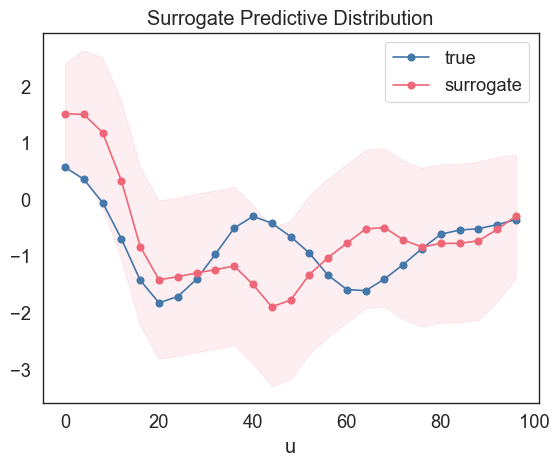}
            \caption{Surrogate Predictive Distribution}
          \label{fig:lin-gauss-surrogate-pred}
        \end{subfigure}
    \end{subfigure}

    \vspace{1em}

    \begin{subfigure}[b]{\textwidth}
        \centering
        \begin{subfigure}[b]{0.45\textwidth}
            \includegraphics[width=\textwidth]{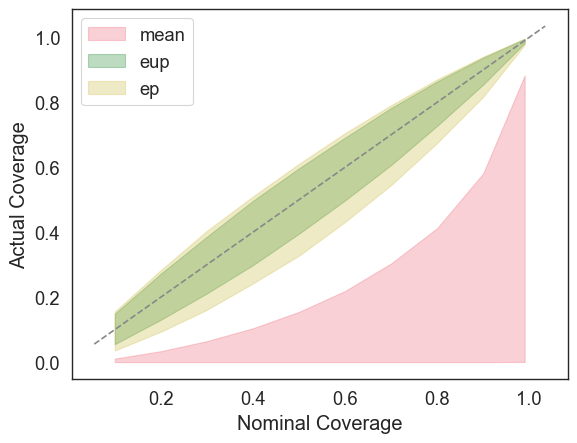}
            \caption{Coverage}
            \label{fig:lin-gauss-coverage}
        \end{subfigure}
        \hfill
        \begin{subfigure}[b]{0.45\textwidth}
            \includegraphics[width=\textwidth]{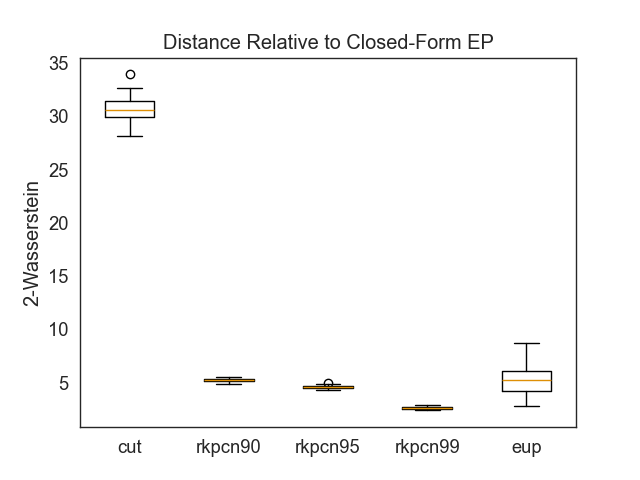}
            \caption{EP Approximation}
            \label{fig:lin-gauss-mcmc}
        \end{subfigure}
    \end{subfigure}

    \caption{
    	\emph{(a)} Summary of the exact posterior for a single replicate: true posterior mean, 
			 \texorpdfstring{$\pm 2$}{plus/minus 2} standard deviations (shaded blue), 
			 \texorpdfstring{$\pm 2$}{plus/minus 2} prior standard deviations, and observations (red).
	\emph{(b)} Marginal distributions of \texorpdfstring{$\targetEm(\Par_{\circ})$,}{} the surrogate evaluated at the ground truth parameter.
			The shaded region encloses \texorpdfstring{$\pm 2$}{plus/minus 2} predictive standard deviations. The surrogate only predicts at the 
			observation locations (points on the plot).
	\emph{(c)} Joint ellipsoidal coverage of the plug-in mean, EUP, and EP approximations, relative to the exact posterior.
			 The shaded regions summarize the middle 90\% of the replicates.
	\emph{(d)} The 2-Wasserstein distance between various approximations to the EP, summarized over the 100 replicates.
			 The distance to EUP is computed in closed-form. The sample-based approximations---independent cut and 
			 RKpCN with \texorpdfstring{$\rho \in \{0, 0.9, 0.95, 0.99\}$}{ρ in (0, 0.9, 0.95, 0.99)}---are fit to Gaussians and then 
			 the closed-form distance formula is applied.
    }
    \label{fig:lin-gauss-combined}
\end{figure}

Finally, we consider a surrogate with $\emCov = G \priorCov G^\top$, 
thus assuming the surrogate bias follows the same smoothness as the underlying signal.
We run one hundred replicate experiments with this setup, with each replicate sampling a 
ground truth parameter $\Par_{\circ}$ from the prior, generating synthetic data using this parameter, and 
sampling the emulator bias $r$ from $\Gaussian(0, \emCov)$. Thus, each surrogate replicate is 
biased but well-calibrated on average. \Cref{fig:lin-gauss-exact-post,fig:lin-gauss-mcmc} display the marginal 
distributions for the exact posterior and surrogate predictive distribution for a single replicate, respectively.

\paragraph{Approximate Posterior Comparison.} \Cref{fig:lin-gauss-coverage} summarizes the joint coverage of
the surrogate-based posteriors relative to the exact baseline. Though the surrogate is well-calibrated, the plug-in mean approximation
severely under-covers due to the unquantified surrogate bias. Both the EUP and EP are well-calibrated and exhibit 
similar behavior. The EP exhibits a marginally larger level of under-coverage for a portion of the replicates. This is not 
surprising given the nature of the surrogate---the surrogate noise essentially acts as another observation noise term,
a problem effectively made for the EUP. The EP nonetheless provides reasonable uncertainty quantification in this example.

\paragraph{MCMC Evaluation.} \Cref{fig:lin-gauss-mcmc} compares the MCMC-based approximations against the 
closed-form EP and EUP distributions. Each of the RKpCN algorithms accurately represent the EP, with the $\rho = 0.99$
setting driving the approximation error near zero. The EUP exhibits more variability, while the independent cut approximation 
consistently deviates from the EP target. These algorithms are operating over a high-dimensional ($\dimPar = 100$)
parameter space in this example. The RKpCN algorithm is well-equipped for this situation, as the pCN proposal is 
designed for high-dimensional latent Gaussian models \citep{functionSpaceMCMC}.

\subsection{Ecosystem Model Calibration} \label{sec:vsem}
Our next example is motivated by the problem of producing near-term forecasts of the 
terrestrial carbon cycle \citep{nearTermForecasts,FerEmulation}. In this setting, model parameters 
are typically unknown and must be learned from observational data. Parameter estimation runs into 
computational challenges for large-scale land surface models, 
underscoring the potential for surrogates in this domain \citep{paramLSM}. 

\subsubsection{Experimental Setup}

\paragraph{Mechanistic Model.}
We consider a synthetic data experiment using the \textit{Very Simple Ecosystem Model} (VSEM), 
a toy model capturing the basic structure of more complex land surface models \citep{vsem}. 
The model, which is an ODE of similar form to that in \Cref{ode_ivp}, is described in detail in
\Cref{app:vsem}.

\paragraph{Statistical Model.}
We consider the task of estimating the parameters $\Par \Def (\alphaV, \vegInit)$, where $\vegInit$ is the initial
condition for the quantity of carbon in the above-ground vegetation pool and $\alphaV$ controls the fraction
of carbon allocated to this pool at each time step. We assume that the observations $\obs$ consist of
noisy monthly averages of leaf area index (LAI), a quantity proportional to the amount of carbon in 
above-ground vegetation. We assume an additive Gaussian noise model
\begin{align}
&\obs = \fwd(\Par) + \noise,
&&\noise \sim \Gaussian(0, \sigma^2 I)
\label{eq:vsem-likelihood}
\end{align}
where $\fwd: \parSpace \to \R^{12}$ maps to the monthly LAI means. For simplicity, we fix $\sigma^2 = 1.0$
and assume independent priors $\alphaV \sim \mathrm{Unif}(0.4, 1.0)$, $\vegInit \sim \mathrm{Unif}(0, 10)$.
Ground truth values $\Par_{\circ}$ are sampled from the prior and $\fwd$ is assumed to be well-specified.
Synthetic data $\obs$ is simulated from the model in \Cref{eq:vsem-likelihood} with $\Par = \Par_{\circ}$.

\paragraph{Surrogate Model.}
We consider two different emulators for the log-posterior $\target(\Par) \Def \log\{\priorDens(\Par) \lik(\Par; \obs)\}$.
The first is a conjugate GP surrogate with GP prior defined by a constant mean function and a Gaussian kernel.
The training data $\{(\Par_n, \target(\Par_n))\}_{n=1}^{\Ndesign}$ is constructed by sampling a set of 
points using a Latin hypercube design and then evaluating the exact log-posterior density at these points. 
The GP mean constant and kernel hyperparameters are optimized by maximum marginal likelihood 
using the \verb+gpjax+ Python package \citep{gpjax}. We constrain the kernel lengthscales to avoid pathologically 
high or low values, and constrain the variance parameter of the GP likelihood to be small to encourage interpolation 
of the training points. Conditioning the optimized GP on the training data yields the surrogate
$\targetEm \sim \GP(\gpMean, \gpKer)$. 

Though this surrogate construction has been used in the literature, it has the apparent deficiency of neglecting 
known bound constraints on the target density. Given the Gaussian likelihood in \Cref{eq:vsem-likelihood}, we
know that at any point $\Par$ the log-posterior cannot exceed $b(\Par) \Def \log\det(2\pi \sigma^2 I) + \log \priorDens(\Par)$.
We thus consider a second surrogate $\targetEm^{\mathrm{clip}}$ defined by the pointwise clipping transform
$\targetEm^{\mathrm{clip}}(\Par) \Def \min\{\targetEm(\Par), b(\Par)\}$. \citet{quantileApprox} make a similar 
observation and instead truncate the predictive distribution, but we find that the clipped Gaussian is more appropriate here.

\paragraph{Experiment Replications.}
We repeat one hundred replications of the above setup, testing both the GP and clipped GP for each replicate.
The replicates vary in the sampled driver data, ground truth $\Par_{\circ}$, synthetic observations $\obs$, 
and design points for the emulator. In addition, the values of 
the VSEM parameters that are not being calibrated are randomly sampled from uniform distributions, yielding
a different parameterization of the forward model for each replicate. This entire procedure is repeated for different 
numbers of design points $\Ndesign \in \{4, 8, 16\}$. As the parameter space here is two-dimensional, 
we approximate all probability distributions and coverage metrics over a dense grid.

\subsubsection{Results}

 \begin{figure}[htbp]
    \centering

    \begin{subfigure}[b]{0.85\linewidth}
        \centering
        \includegraphics[width=\linewidth]{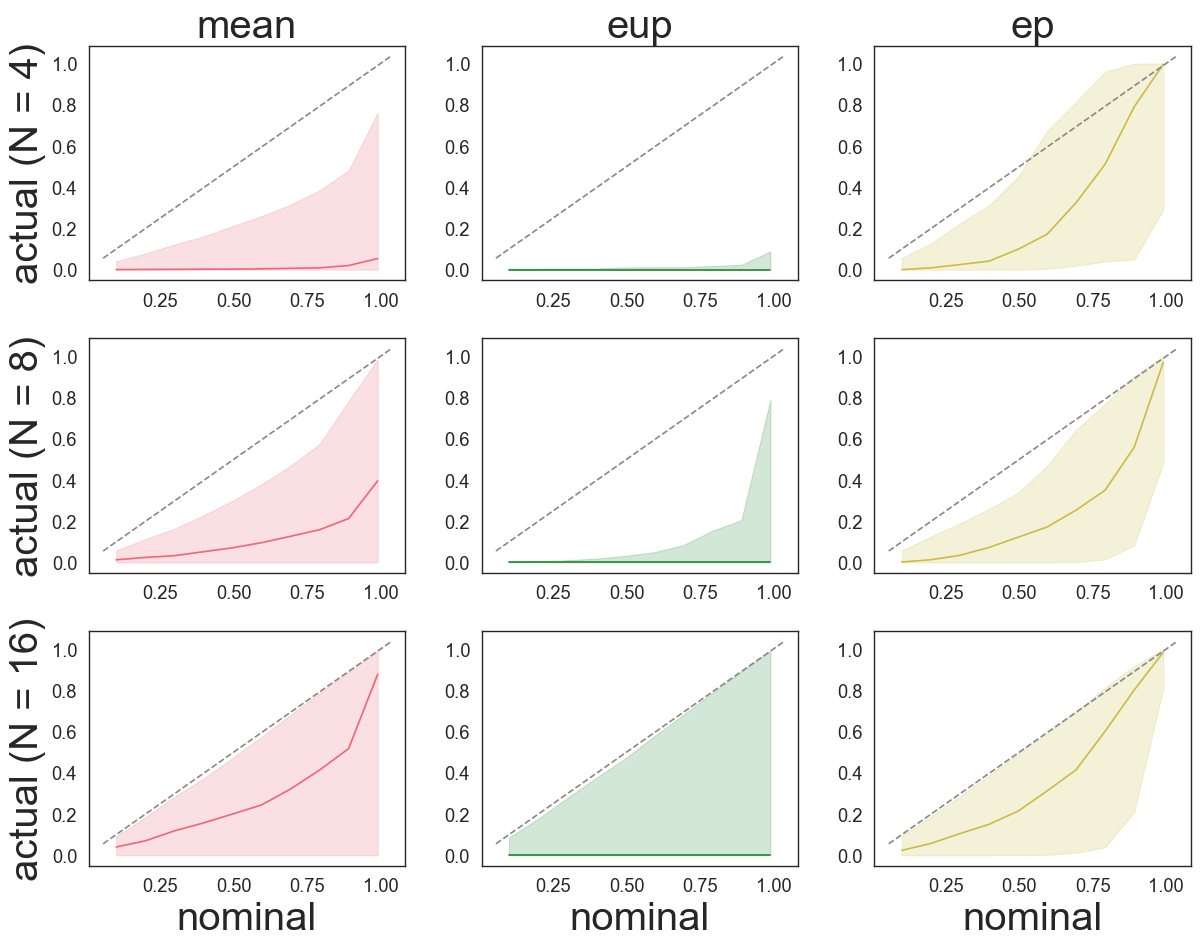}
        \caption{Coverage results using \texorpdfstring{$\targetEm$}{GP}}
        \label{fig:vsem_coverage_gp}
    \end{subfigure}
    
    \begin{subfigure}[b]{0.85\linewidth}
        \centering
        \includegraphics[width=\linewidth]{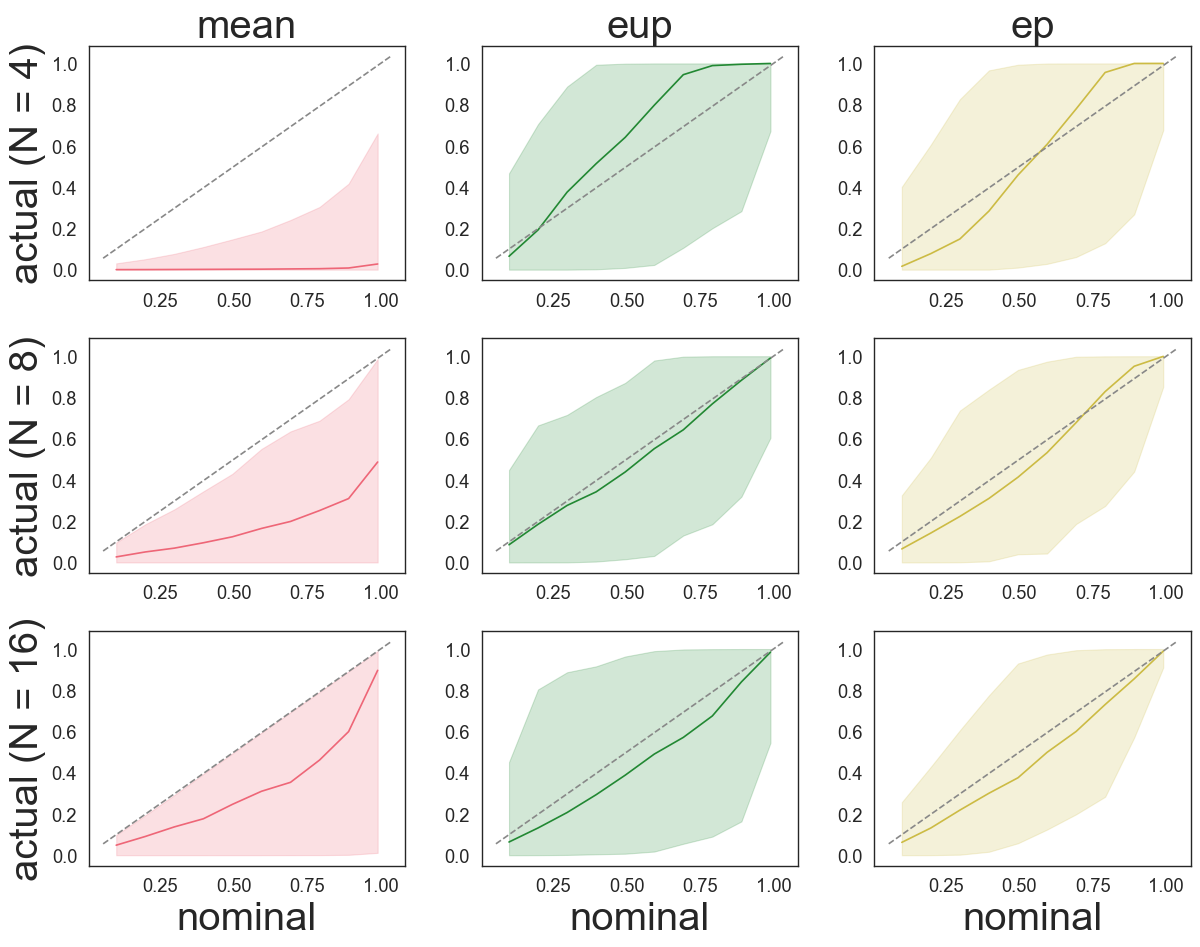}
        \caption{Coverage results using \texorpdfstring{$\targetEm^{\mathrm{clip}}$}{clipped GP}}
        \label{fig:vsem_coverage_clip_gp}
    \end{subfigure}

    \caption{
        Coverage metrics for the surrogate-based posteriors in the VSEM experiment, using the
    	GP emulator \texorpdfstring{$\targetEm$}{} \emph{(top)} and clipped GP emulator 
	$\texorpdfstring{\targetEm^{\mathrm{clip}}}{}$ \emph{(bottom)}. Coverage is computed using masks over a
	dense two-dimensional grid. The shaded regions summarize the middle 90\% of the replicates.
    }
    \label{fig:combined-vsem-coverage}
\end{figure}

\paragraph{GP Surrogate.}
\Cref{fig:vsem_coverage_gp} summarizes the coverage of the posterior approximations derived using the GP 
emulator $\targetEm$. The mean and EUP approximations systematically under-cover, while the EP is better 
calibrated, though still under-covers on average. Very few replicates for any of the approximations actually
over-cover, as would be expected for the approximations that propagate surrogate uncertainty. This is largely 
due to the extreme sensitivity of posterior approximations derived from GP log-density surrogates.
As noted in \Cref{ex:ldens-em-cont}, the EUP tends to concentrate in the region with highest uncertainty. When 
there are fewer design points then the GP predictive variance is larger and this concentration can be extreme, 
as evidenced by the severe under-coverage in the $\Ndesign = 4$ case. Even in the $\Ndesign = 16$ case,
in which the GP variance is typically modestly small, the median EUP replicate has a coverage of zero at 
every probability level. The GP model itself bears partial responsibility---it is generally quite difficult to 
encode reasonable inductive biases in a GP model to accurately represent the epistemic uncertainty for a complex 
log-density surface. That being said, the EUP tends to amplify any pathologies of the GP model, while the EP 
is seen to largely avoid the most extreme behavior. In general, we urge caution in using a 
global GP log-density emulator without the incorporation of additional inductive biases to guard against
this sensitive behavior. 

\paragraph{Clipped GP Surrogate.}
The clipped GP results in \Cref{fig:vsem_coverage_clip_gp} demonstrate that the small adjustment of encoding the 
likelihood bound constraint in the predictive distribution drastically improves all of the posterior approximations.
The plug-in mean still systematically under-covers, but both the EUP and EP are well-calibrated on average.
The clipping transform avoids the EUP pathologies stemming form the log-normal tails of $\postEm(\Par)$.
However, we do still observe evidence (especially for $\Ndesign = 16$) of EUP instability relative to the EP,
with a greater portion of the replicates resulting in significant under or over-coverage.

\subsection{Spatial PDE-Constrained Inversion} \label{sec:pde}
We next consider PDE-constrained estimation of a latent spatial field. This experiment is motivated by hydrological
applications in which a heterogeneous permeability field must be estimated from pressure observations at a set of 
spatial locations \citep{conditionPermToPressure,mcmcHydrology,precondMCMCKL}.
Challenges of such inverse problems include the infinite dimensionality of the parameter and the high cost of
the forward model \citep{dimRedPolyChaos}.

\subsubsection{Experimental Setup}
\paragraph{Mechanistic Model.}
Assume that the permeability $\perm(\state)$ and pressure $\pressure(\state)$ fields are related by the elliptic PDE
\begin{equation}
\frac{\partial}{\partial \state} \left\{\perm(\state)\frac{\partial\pressure(\state)}{\partial \state} \right\} = -\source(\state),
\qquad \state \in [0, 1]
\end{equation}
subject to the boundary conditions $\pressure(1) = 1$ and $\perm(0)\frac{\partial\pressure}{\partial \state}(0) = 1$.
The source term is defined as 
\begin{equation}
\source(\state) = \sum_{i=1}^{4} \frac{0.8}{\delta\sqrt{2\pi}} \exp\left\{-\frac{1}{2\delta^2}(\state - c_i)^2 \right\},
\end{equation}
with $\delta = 0.05$ and $(c_1, \dots, c_4) = (0.2, 0.4, 0.6, 0.8)$. We discretize the spatial dimension over an 
evenly-spaced grid $X = \{\state_m\}_{m=1}^{100}$ and solve the PDE with 
a finite difference scheme. Henceforth, we write $\pressure(X) \in \R^{100}$ to denote the output of the 
solver over the grid, given the discretized input field $\perm(X) \in \R^{100}$.

\paragraph{Statistical Model.}
We assume that noisy pressure measurements are obtained at four of the grid points 
$X^{\mathrm{obs}} = \{\stateObs_1, \dots, \stateObs_4\} \subset X$. 
The goal is to recover the permeability field given these measurements. We assume the observation model 
$\obs = \pressure(X^{\mathrm{obs}}) + \noise$, $\noise \sim \Gaussian(0, \sigma^2 I)$ with 
$\sigma = 0.001$. The log-permeability field is given a GP prior $\log \perm \sim \GP(1, k_\kappa)$
with a constant mean of one and an exponential kernel. This implies the Gaussian prior 
$\log \perm(X) \sim \Gaussian(1, k_\kappa(X, X))$ for the discretized field. 
To reduce the parameter dimension, we approximate this Gaussian prior by retaining only the dominant six
principal components of $k_\kappa(X, X)$. This tends to capture upwards of 95\% of the prior variance under 
our experimental setup, and mimics the common approach of invoking Karhunen-Loève approximations in 
spatial inverse problems \citep{uribeKL,dimRedPolyChaos,KLInvProbs}. Letting $\{(\lambda_r, \psi_r)\}_{r=1}^{6}$ denote 
the dominant eigenpairs, the rank-reduced prior model is
\begin{equation}
\log \perm(X) \Def 1 + \sum_{r=1}^{6} \sqrt{\lambda_r} \Par_j \psi_j, \qquad \Par_j \overset{\mathrm{iid}}{\sim} \Gaussian(0, 1),
\end{equation}
so that the six parameters $\Par \Def (\Par_1, \dots, \Par_6)^\top$ now control the permeability field.
Letting $\fwd: \R^{6} \to \R^{4}$ denote the map from $\Par$ to 
$\pressure(X^{\mathrm{obs}}) \in \R^{4}$, the final inverse problem can be written as
\begin{equation}
\obs \given \Par \sim \Gaussian(\fwd(\Par), \sigma^2 I), \qquad \Par \sim \Gaussian(0, I).
\label{eq:pde-inv-prob}
\end{equation}
While the effects of spatial discretization and prior approximation are of interest, we neglect such questions 
and treat \Cref{eq:pde-inv-prob} as the baseline ``exact'' model in order to focus on questions of surrogate uncertainty propagation.
Ground truth values $\Par_{\circ}$ are sampled from the prior, and synthetic data is generated from the model in 
\Cref{eq:pde-inv-prob} with $\Par = \Par_{\circ}$.

\paragraph{Surrogate Model.}
We fit a multi-output GP to the forward model $\target \Def \fwd$, consisting of independent single-output GPs fit to each of
the four scalar outputs. The training data $\{(\Par_n, \target(\Par_n))\}_{n=1}^{\Ndesign}$ is constructed by 
mapping Latin hypercube samples through the standard Gaussian 
quantile function to obtain the design inputs, and then evaluating $\fwd$ at each design input.
Each univariate GP is defined by a constant mean function and a Gaussian kernel, with hyperparameters
optimized by maximum marginal likelihood using \verb+gpjax+ \citep{gpjax}.
Conditioning the optimized GP on the training data yields the multi-output forward model surrogate
$\targetEm \sim \GP(\gpMean, \gpKer)$. For all of the surrogate-based posterior approximations, we 
truncate the prior support to $[-\alpha, \alpha]^6$, where $\alpha$ is the 99.9\% percentile of the 
standard normal distribution. This captures about 99\% of the prior mass, and is done to avoid pathological 
behavior stemming from the surrogate's reversion to the GP prior in the tails. Similar practical tactics are
employed in \citet{gp_surrogates_random_exploration,FerEmulation} when fitting surrogates over 
an unbounded support.

\paragraph{Experiment Replications.}
We repeat one hundred replications of the above setup. Each replicate varies in the sampled ground 
truth $\Par_{\circ}$, synthetic observations $\obs$, and design points for the emulator. This entire procedure 
is repeated for different numbers of design points $\Ndesign \in \{10, 20, 30\}$. Inference for the exact, 
plug-in mean, and EUP distributions is conducted via Metropolis-Hastings, retaining 5000 samples 
each after dropping burn-in and thinning. Approximate EP samples are obtained using RKpCN 
(\Cref{alg:rk-pcn-infinite}). It is difficult to obtain an exact EP baseline in this higher dimensional example
with an infinite-dimensional surrogate. We construct an EP baseline by first obtaining a finite-dimensional 
GP approximation using the pathwise sampling approach described in 
\citet{pathwiseConditioning,samplingGPPosts}, with the GP prior approximated using 1000 random 
Fourier features \citep{RFF}. Preliminary experiments showed that this approximation was typically very
accurate. Using the resulting finite-dimensional basis, we run a Monte Carlo within Metropolis scheme;
in particular, we sample 100 surrogate replicates, run 
Metropolis-Hastings on the induced posterior trajectories, and retain 10 samples from each trajectory.
The choice of 100 replicates was chosen to keep computation time reasonable; this may be inadequate
for a complete characterization of the EP, so should not be viewed as an exact baseline in this experiment.

\subsubsection{Results}

\begin{figure}[htbp]
    \centering
    \includegraphics[width=\linewidth]{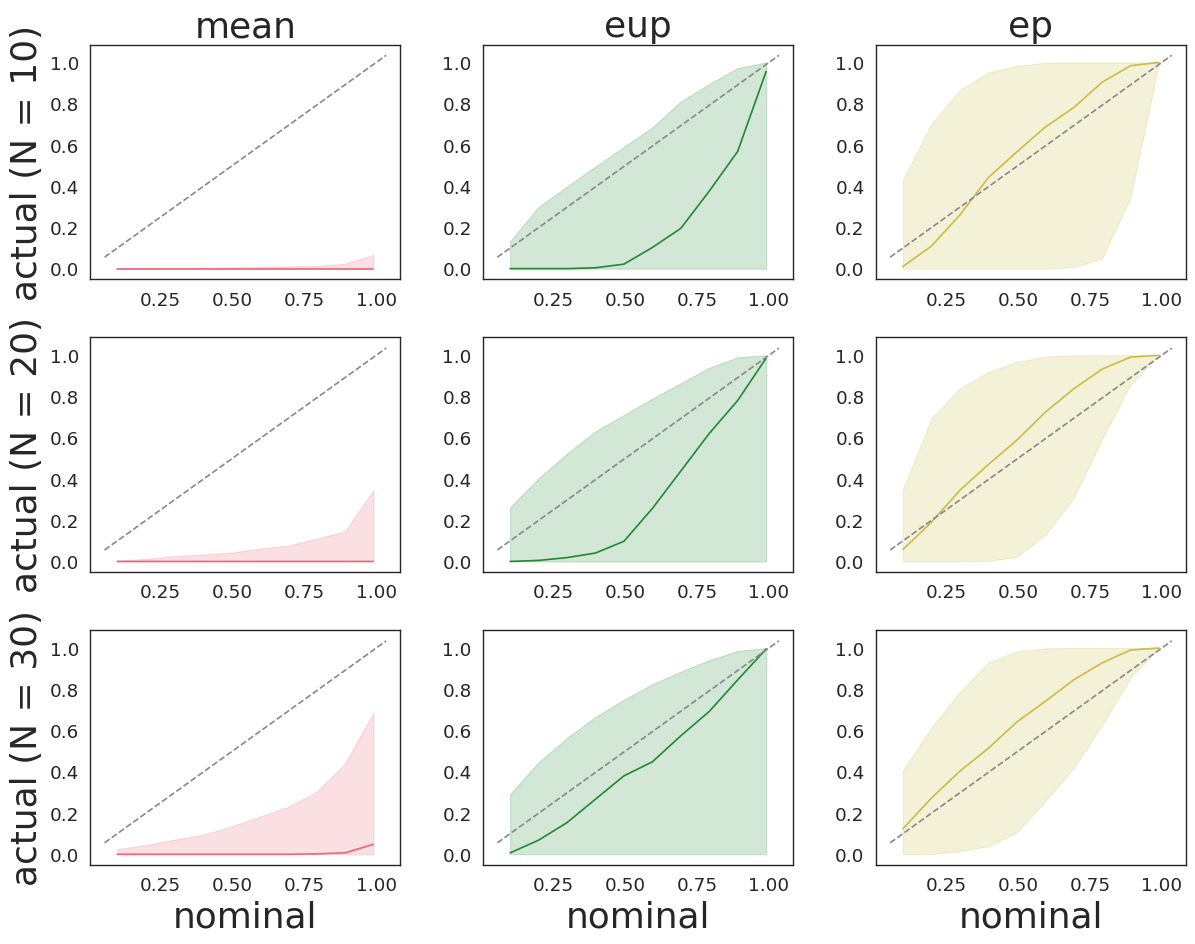}
    \caption{Coverage results for the surrogate-based posteriors in the PDE experiment.
    		 Computed by estimating ellipsoidal coverage regions based on Mahalanobis distance using MCMC samples.
		 The shaded regions summarize the middle 90\% of the replicates.}
    \label{fig:pde_coverage}
\end{figure}

\begin{figure}[htbp]
    \centering
    \begin{subfigure}[b]{0.32\textwidth}
        \includegraphics[width=\textwidth]{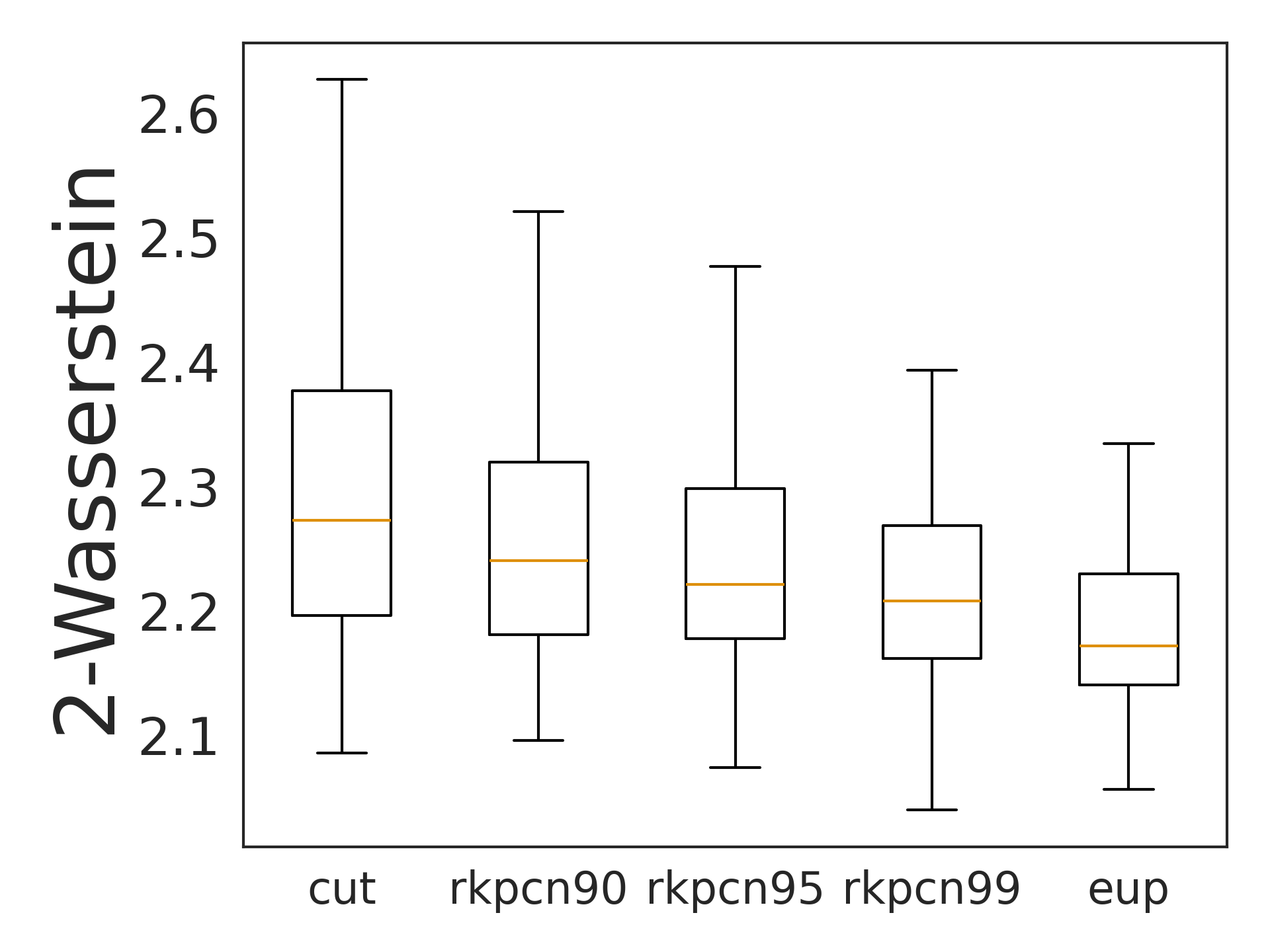}
        \caption{\texorpdfstring{$N = 10$}{N = 10}}
    \end{subfigure}
    \hfill
    \begin{subfigure}[b]{0.32\textwidth}
        \includegraphics[width=\textwidth]{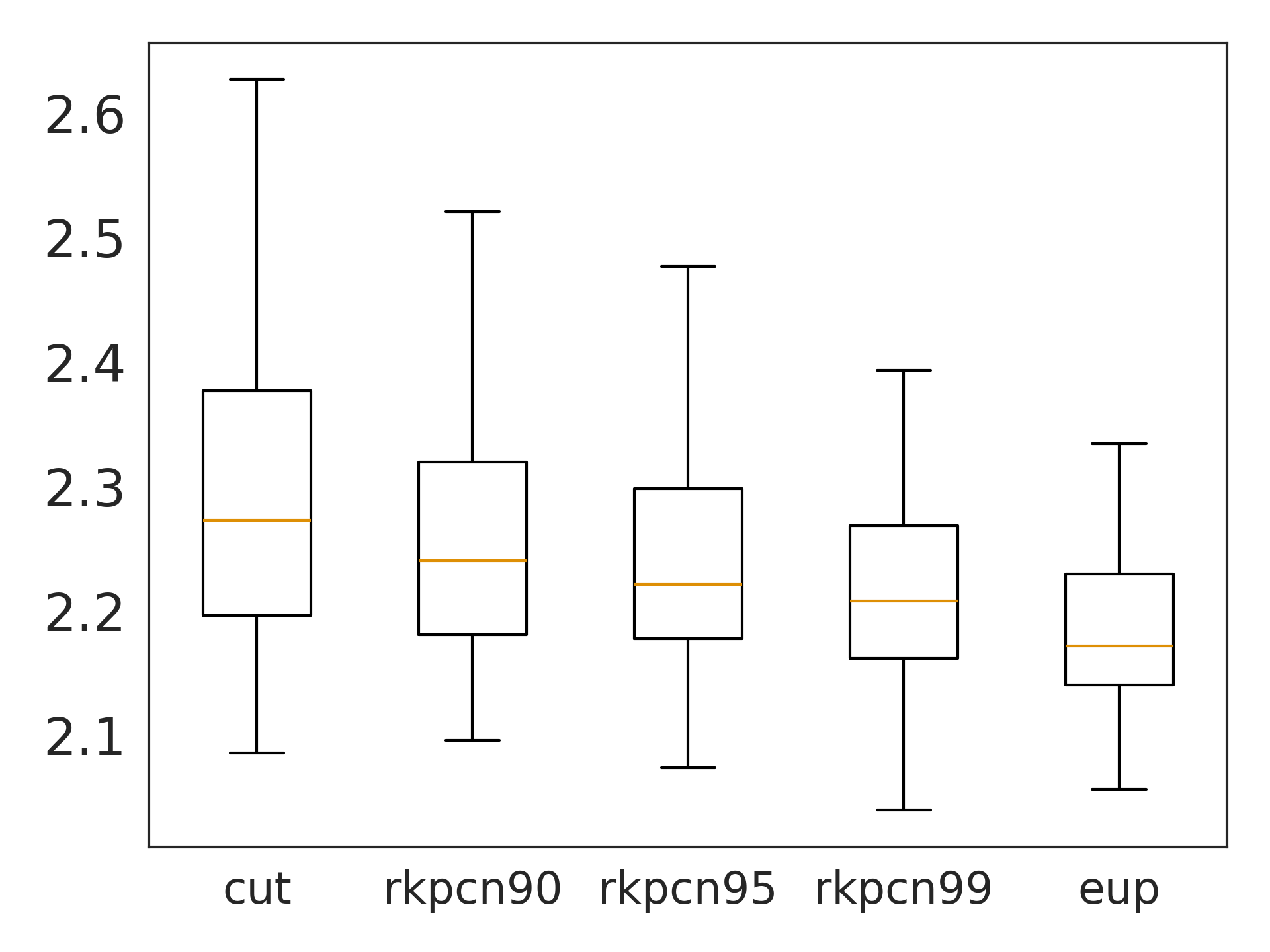}
        \caption{\texorpdfstring{$N = 20$}{N = 20}}
    \end{subfigure}
    \hfill
    \begin{subfigure}[b]{0.32\textwidth}
        \includegraphics[width=\textwidth]{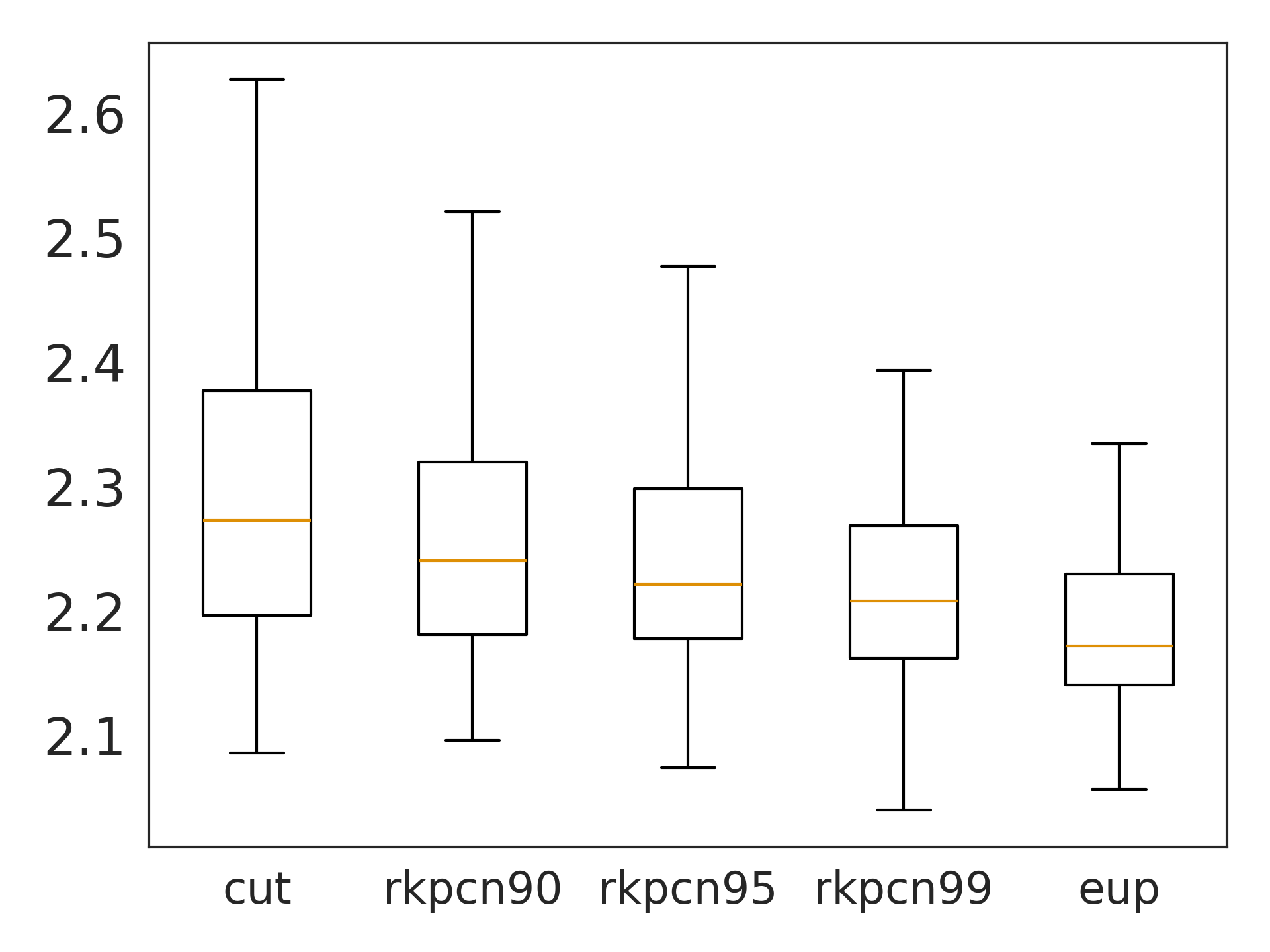}
        \caption{\texorpdfstring{$N = 30$}{N = 30}}
    \end{subfigure}
    \caption{EP approximations in the PDE experiments.}
    \label{fig:pde_wasserstein}
\end{figure}

\paragraph{Approximate Posterior Comparison.} \Cref{fig:pde_coverage} summarizes the joint ellipsoidal coverage of
the surrogate-based posteriors relative to the exact baseline. Ellipsoidal coverage (based only on the empirical mean and 
covariance of MCMC samples from the approximating distributions) was deemed reasonable upon observing that 
the samples tended to have roughly elliptical shape across experimental replicates. The EP is seen to be well-calibrated on
average across the three design sizes, while the plug-in mean significantly under-covers. The EUP tends to under-cover, 
but the median replicate is nearly well-calibrated at the largest design size. Even at this design size, a significant portion of 
the EUP replicates continues to completely miss the support of the true posterior. Naturally, the calibration of the posteriors
can always be improved by improving the calibration of the underlying surrogate. We sought to fit reasonably well-calibrated
GPs with realistic degrees of bias.

\paragraph{MCMC Evaluation.} \Cref{fig:pde_wasserstein} summarizes the distance between various sample-based
approximations to the MwMC EP baseline. We use entropic-regularized 2-Wasserstein distance,
computed using the Sinkhorn algorithm in \verb+ott-jax+ \citep{ottjax}. Each distribution is represented by 1000 sub-samples
from an MCMC algorithm. A whitening transformation is applied to the samples using the empirical mean and covariance of 
the MwMC so that the distances are computed in the baseline MwMC geometry. The Sinkhorn regularization parameter is 
selected using the \verb+ott-jax+ default computed with respect the MwMC points, and the same value is used for all 
Wasserstein computations for a fixed design size. The results show that both RKpCN with $\rho = 0.99$ and the EUP 
provide the best EP approximations, with the median EUP replicate performing slightly better. 

\section{Discussion and Conclusions} \label{sec:conclusion}
We provide theoretical arguments that generally favor the EP as the correct target for surrogate-based Bayesian inference.
We compare this baseline to the EUP, a computationally-convenient alternative, and demonstrate that the two distributions
can significantly deviate when the surrogate-based posterior density approximation has highly non-uniform
uncertainty over the parameter space. In particular, this occurs when the normalizing constant $\normCst(\targetEm)$ is highly 
correlated with $\targetEm(\Par)$ in small regions of the parameter space.
This problem is exemplified by applications where GPs are used to emulate 
log-densities, but can sometimes be mitigated by practical safeguards (e.g., enforcing bound constraints). In our experiments,
the EUP and EP tend to exhibit closer agreement in the forward model emulation setting. This is likely due to the fact
that our numerical experiments consider Gaussian likelihoods, which are bounded and thus dampen the effects of high 
surrogate uncertainty in small regions of the parameter space. 
Caution is warranted when performing surrogate-based Bayesian inference with unbounded likelihoods.
We also highlight different computational strategies for EP-based inference, and present the RKpCN algorithm, an easy-to-implement 
approximate EP sampler that naturally handles the infinite-dimensionality of GP surrogates.

There are several avenues for future research. First, it may be beneficial to explore alternative conceptual frameworks
other than the Bayesian decision-theoretic framework considered here. One option is to consider regularizing the
optimization problem in \Cref{prop:EP-variational} to yield an EP-like target that exhibits prior-reversion when surrogate
uncertainty is high. This would naturally lead to connections with generalized Bayesian inference 
(e.g., \citet{generalizedCut}). Expanding on our discussion of connections with cut posterior inference could also be 
interesting. While the EP can be viewed as a cut posterior, surrogate-based uncertainty propagation 
problems often have specific structure that differs from typical examples considered in the modular Bayes
literature; e.g., the surrogate (the parameter in the first stage of the cut model) may be high or infinite
dimensional, can be directly sampled, and may exhibit significant variation in its conditional distributions
(i.e., surrogate variation over the parameter space). As a starting point, we considered experiments with 
well-specified inverse problem likelihoods. Given that the cut posterior is designed for joint distributions 
in which the second-stage model may be misspecified, it would be interesting to investigate whether the benefits of 
the EP over the EUP become more pronounced under the presence of likelihood misspecification.
Conversely, it may also be fruitful to investigate the effects of misspecification in the (first-stage) surrogate
model. If the surrogate correlation structure over the parameter space is highly misspecified, then the EUP 
(which ignores predictive correlations) may be more robust in such cases.

Finally, there is significant potential to further refine the RKpCN algorithm. We justify the algorithm with 
heuristic and empirical justifications in this work to demonstrate that a simple approximation can provide 
a reasonable characterization of the EP. Rigorous theoretical analysis of this algorithm would likely draw
upon results from both the noisy MCMC \citep{noisyMCSurvey,stabilityNoisyMH,corrPM} and function-space MCMC
\citep{functionSpaceMCMC} literatures. There is likely room for improving the naive normalizing constant ratio 
estimate used in RKpCN while maintaining a practical, easily implemented algorithm.

\appendix

\section{Proofs}
For the below proofs we use the following measure-theoretic setup.
Let $\refMeas$ be a reference measure (e.g., Lebesgue) on $(\parSpace, \BorelSig)$,
and $\emDist$ a probability measure on $(\emSpace, \emSig)$. Assume that 
the map $(\Par, \targetTraj) \mapsto \postDens(\Par; \targetTraj)$ from $\parSpace \times \emSpace$
to $\R_{\geq 0}$ is measurable
and $\normCst(\targetTraj) \Def \int \postDens(\Par; \targetTraj) \emDist(\d\targetTraj) \in (0,\infty)$
$\emDist$-almost surely. Let $\postDensNorm(\Par; \targetTraj) \Def \postDens(\Par; \targetTraj) / \normCst(\targetTraj)$
and define the joint distribution 
$\emJoint(\d\Par, \d\targetTraj) \Def \postDensNorm(\Par; \targetTraj) \refMeas(\d\Par)\emDist(\d\targetTraj)$.
Define $\postApproxEP$ to be the density corresponding to the $\Par$-marginal of $\emJoint$; that is,
$\postApproxEP(\Par) \Def \int \postDensNorm(\Par; \targetTraj) \emDist(\d\targetTraj)$. 

\subsection{Proof of \texorpdfstring{\Cref{prop:EP-variational}}{EP Variational Result}}
\paragraph{KL Divergence.} We start by proving the KL divergence result.
Let $\qMeas$ be a probability measure on $\parSpace$ with 
$\refMeas$-density $\qDens$. We restrict to measures 
$\postApproxEP \ll \qMeas$, as the KL divergence is infinite otherwise. 
Note that 
$\frac{\d\emJoint}{\d(\emDist \otimes \qMeas)}(\Par, \targetTraj) = \frac{\postDensNorm(\Par; \targetTraj)}{\qDens(\Par)}$.
Using Tonelli's theorem, we have 
\begin{align*}
\E_{\emDist}\left[\KL{\postNormEm}{\qMeas} \right]
&= \int_{\emSpace} \int_{\parSpace} 
\postDensNorm(\Par; \targetTraj) \log \frac{\postDensNorm(\Par; \targetTraj)}{\qDens(\Par)}
\refMeas(\d\Par) \emDist(\d\targetTraj) \\
&= \int_{\parSpace \times \emSpace}  \log \frac{\postDensNorm(\Par; \targetTraj)}{\qDens(\Par)}
\emJoint(\d\Par, \d\targetTraj) \\
&= \int_{\parSpace \times \emSpace}  \log \frac{\d\emJoint}{\d(\emDist \otimes \qMeas)}(\Par, \targetTraj)
\ \emJoint(\d\Par, \d\targetTraj) \\
&= \KL{\emJoint}{\emDist \otimes \qMeas}.
\end{align*}
Finally, 
\begin{align*}
\KL{\emJoint}{\emDist \otimes \qMeas}
&= \int \log \frac{\d\emJoint}{\d(\emDist \otimes \qMeas)} \d\emJoint \\
&= \int \log \left[\frac{\d\emJoint}{\d(\emDist \otimes \postApproxEP)} \frac{\d(\emDist \otimes \postApproxEP)}{\d(\emDist \otimes \qMeas)}\right] d\emJoint \\
&\proptoAdd \int \log \frac{\d(\emDist \otimes \postApproxEP)}{\d(\emDist \otimes \qMeas)} \d\emJoint \\
&= \int_{\emSpace \times \parSpace} 
\log \frac{\postApproxEP(\Par)}{\qDens(\Par)} \postDensNorm(\Par; \targetTraj) \emDist(\d\targetTraj) \refMeas(\d\Par) \\
&= \int_{\parSpace} \log \frac{\postApproxEP(\Par)}{\qDens(\Par)} 
 \postApproxEP(\Par) \refMeas(\d\Par) \\
&= \KL{\postApproxEP}{\qMeas},
\end{align*} 
where we have used $\proptoAdd$ to absorb additive constants with respect to $\qMeas$. The result follows 
from the fact that $\KL{\postApproxEP}{\qMeas}$ is uniquely minimized at $\qMeas = \postApproxEP$. $\qquad \blacksquare$

\paragraph{Squared \texorpdfstring{$L_2$}{L2} loss.} 
For the expected squared error objective, apply Tonneli's theorem 
\begin{align}
\E_{\emDist}\left[\norm{\postNormEm - \qMeas}^2_{L_2(\parSpace)} \right]
&= \int \E_{\emDist} \left[\postDensNorm(\Par; \targetTraj) - \qDens(\Par) \right]^2 \refMeas(d\Par) \nonumber
\end{align}
and notice that the integrand is minimized pointwise by
$\qDens(\Par) = \E_{\emDist}[\postDensNorm(\Par; \targetTraj)] = \postApproxEP(\Par)$. $\qquad \blacksquare$
 
\subsection{Proof of \texorpdfstring{\Cref{prop:kl-cut-op}}{Cut Posterior Result}}
Recall the joint distribution 
$\jointKOH(\d\Par, \d\targetTraj, \d\obs, \d\emObs) =
\priorDens(\Par) \lik(\Par; \targetTraj, \obs) \emLik(\targetTraj; \emObs) 
\emDistPrior(\d\targetTraj) \, \d\Par \, \d\obs \, \d\emObs$
with conditional $\postKOH(\d\Par, \d\targetTraj)$.
Let $\postKOH_{\targetTraj}(\d\Par)$ denote the $\targetTraj$-marginal of this conditional.
Similarly, let $\condMargKOH$ denote the marginal conditional of 
$\Par$ given $(\obs,\targetTraj)$.
By the disintegration theorem, any $\qMeas \in \qSpaceCut$ can be written as 
$\qMeas(\d\Par, \d\targetTraj) = \emDist(\d\targetTraj) \qCond(\targetTraj, \d\Par)$ since 
$\qSpaceCut$ restricts the $\targetTraj$-marginal of $\qMeas$ to equal $\emDist$.
Subject to standard regularity conditions, it follows that
\begin{equation}
\frac{\d\qMeas}{\d\postKOH}(\Par, \targetTraj) =
\frac{\d\emDist}{\postKOH_{\targetTraj}(\d\Par)}(\targetTraj)
\frac{\d\qCond(\targetTraj,\cdot)}{d\condMargKOH}(\Par).
\end{equation}
Therefore,
\begin{align*}
\KL{\qMeas}{\postKOH} 
&= \int \log \left[\frac{\d\qMeas}{\d\postKOH}\right] \qMeas(\d\Par, \d\targetTraj) \\
&\proptoAdd \int \log \left[\frac{\d\qCond(\targetTraj,\cdot)}{d\condMargKOH}(\Par)\right] \emDist(\d\targetTraj) \qCond(\targetTraj,\d\Par) \\
&= \int_{\emSpace} \left\{\int_{\parSpace}  
\log \left[\frac{\d\qCond(\targetTraj,\cdot)}{d\condMargKOH}(\Par)\right] \qCond(\targetTraj,\d\Par) \right\} \emDist(\d\targetTraj) \\
&= \int_{\emSpace} \KL{\qCond(\targetTraj,\cdot)}{\condMargKOH} \emDist(\d\targetTraj).
\end{align*}
Since the integrand is minimized pointwise by $\qCond(\targetTraj,\cdot) = \condMargKOH$, it follows that
$\qMeasOpt(\d\Par, \d\targetTraj) = \emDist(\d\targetTraj)\condMargKOH(\d\Par)$. $\qquad \blacksquare$
 
\subsection{Proofs of \texorpdfstring{\Cref{prop:ep-eup-pw-err}}{EUP/EP pointwise error} and \texorpdfstring{\Cref{prop:ep-eup-TV-err}}{Average Error}}
Recall that for two random variables $a$ and $b$ it holds that $\E[ab] = \E[a]\E[b] + \Cov(a,b)$.
Applying this identity, we have
\begin{equation*}
\postApproxEP(\Par)
= \emE[\postEm(\Par) \normCstEm^{-1}]
= \emE[\postEm(\Par)] \emE[\normCstEm^{-1}] + \Cov(\postEm(\Par), \normCstEm^{-1}).
\end{equation*} 
Subtracting $\postApproxEUPNorm(\Par) = \emE[\postEm(\Par)] / \emE[\normCstEm]$
and grouping terms completes the derivation. The integrated error follows from applying the 
triangle inequality, integrating over $\parSpace$, and applying Tonelli's theorem, which gives
\begin{equation*}
\int \emE[\postEm(\Par)] \refMeas(\d\Par) = \emE \int \postEm(\Par) \refMeas(\d\Par) = \emE[\normCstEm]
\qquad \blacksquare
\end{equation*} 

\section{MCMC Algorithm Details} \label{app:mcmc}
The below algorithm provides a practical implementation of \Cref{alg:rk-pcn-finite} that only requires the realization
of finite dimensional projections of the surrogate trajectories. This relies on the fact that $\postDens(\Par; \targetTraj)$
is a function of $\targetTraj$ only through $\targetTraj(\Par)$. The algorithm requires one just-in-time GP sample 
each iteration. However, additional conditioning points are not accumulated across iterations, avoiding the numerical
issues associated with repeated just-in-time sampling required by MwMC methods (\Cref{alg:ep}) or the tempered 
transitions strategy of \citet{PlummerCut}.

\begin{algorithm}[H]
    \caption{Random Kernel pCN, Practical Implementation  (single iteration)}
    \label{alg:rk-pcn-infinite}
    \begin{algorithmic}[1]
    \State \textbf{Input:} Current state $(\Par, \targetTraj_{\Par})$
    \State \textbf{Output:} Updated state $(\nextstate{\Par}, \nextstate{\targetTraj}_{\nextstate{\Par}})$
    \State $\propPar \sim \propDistPar(\Par, \cdot), v_\Par \sim \mathrm{Unif}(0,1)$
    \State $\targetTraj_{\propPar} \sim \mathrm{law}(\targetEm(\propPar) \given \targetEm(\Par) = \targetTraj_{\Par})$ \Comment{Just-in-time sample}
    \State $\gpMean^{\Par, \propPar} \Def (\gpMean(\Par), \gpMean(\propPar))^\top$, $K_{\star}^{\Par, \propPar} \Def \gpKer((\Par, \propPar), (\Par, \propPar))$ \Comment{Bivariate projection}
    \State $\xi \sim \Gaussian(0, K_{\star}^{\Par, \propPar})$
    \State $(\nextstate{\targetTraj}_{\Par}, \nextstate{\targetTraj}_{\propPar})^\top \gets 
    	\gpMean^{\Par, \propPar} + \pcnCor \{(\targetTraj_{\Par}, \targetTraj_{\propPar})^\top  - \gpMean^{\Par, \propPar}\} + \sqrt{1 - \pcnCor^2} \xi$ \Comment{$\targetTraj$ update}
    \State $\alpha_{\Par} \gets \min\left\{1, \postDens(\propPar; \nextstate{\targetTraj}) / \postDens(\Par; \nextstate{\targetTraj}) \right\}$ 
    	\Comment{Compute using $\nextstate{\targetTraj}_{\Par}, \nextstate{\targetTraj}_{\propPar}$}
    \State $\nextstate{\Par} \gets \propPar \indicator{v \leq \alpha_{\Par}} + \Par \indicator{v > \alpha_{\Par}}$ \Comment{$\Par$ update}
    \State $\nextstate{\targetTraj}_{\nextstate{\Par}} \gets \nextstate{\targetTraj}_{\propPar} \indicator{v_\Par \leq \alpha_{\Par}} + \nextstate{\targetTraj}_{\Par} \indicator{v_\Par > \alpha_{\Par}}$
    \end{algorithmic}
\end{algorithm}

\section{Additional Details for Numerical Experiments}

\subsection{Linear Gaussian Example}
\begin{figure}[H]
    \centering
    \begin{subfigure}{0.7\linewidth}
        \centering
        \includegraphics[width=\linewidth]{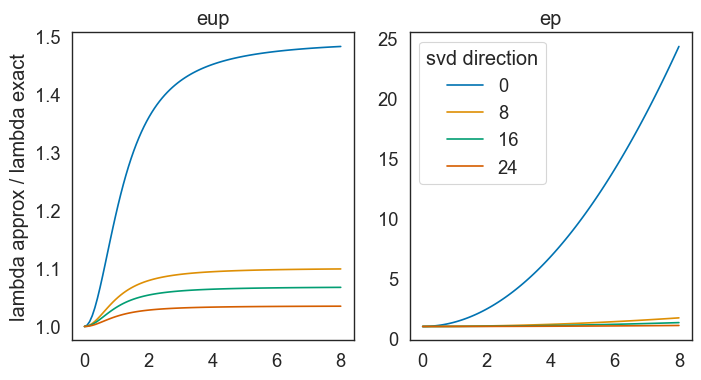}
    \end{subfigure}

    \begin{subfigure}{0.7\linewidth}
        \centering
        \includegraphics[width=\linewidth]{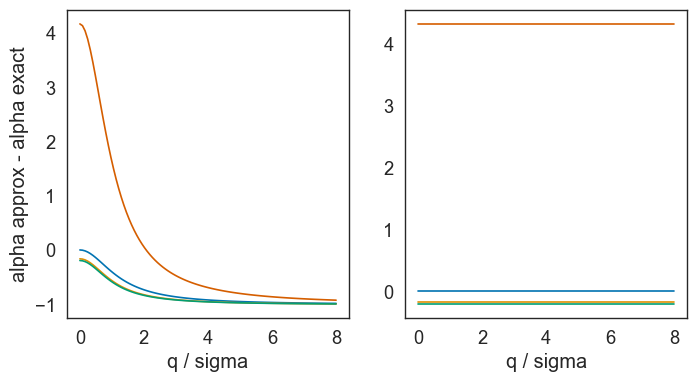}
    \end{subfigure}
    \vspace{0.5em}

    \caption{Scaling behavior as a function of \texorpdfstring{$q / \sigma$}{q / σ}, with 
    \texorpdfstring{$c_0 = 1.0$}{c_0 = 1.0}, \texorpdfstring{$\sigma=1.0$}{σ = 1.0}, \texorpdfstring{$r = (2.5, \dots, 2.5)^\top}{r = (2.5, ..., 2.5)}$.
    The forward model \texorpdfstring{$G$}{G} is the convolution operator used in \Cref{sec:deconvolution}, with parameter and 
    data dimensions $\texorpdfstring{\dimPar}{D} = 100$, $\texorpdfstring{\dimObs}{P} = 25$.
    The plots show the variance and mean coordinates along certain \texorpdfstring{$v_j$}{v_j} directions 
    (smaller \texorpdfstring{$j$}{j} indicating directions better informed by data),
    relative to the exact (no surrogate) posterior analogs. Starting from the top-left in clockwise
    order, the plots display \texorpdfstring{$\lambda_j^{\eup} / \lambda_j$}{λ_j^eup / λ_j}, 
    \texorpdfstring{$\lambda_j^{\ep} / \lambda_j$}{λ_j^ep / λ_j}, 
    \texorpdfstring{$\alpha_j^{\eup} - \alpha_j$}{α_j^eup - α_j}, and 
    \texorpdfstring{$\alpha_j^{\ep} - \alpha_j$}{α_j^ep - α_j} as a function of \texorpdfstring{$q / \sigma$}{q / σ}.}
    \label{fig:lin-gauss-svd-plots}
\end{figure}

\Cref{fig:lin-gauss-svd-plots} provides a visualization of the analytical results from 
\Cref{prop:linear-gaussian-analytical}, using the linear forward model from the 
example in \Cref{sec:deconvolution}.

\subsection{Ecological Model Example} \label{app:vsem}
\paragraph{Details for VSEM Model.}
The model 
describes the evolution of the state vector 
$\state(\Time) \Def [\stateV(\Time), \stateR(\Time), \stateS(\Time)]^\top \in \R_{\geq 0}^{3}$,
with the state variables representing the quantity of carbon (\textrm{kg C/$m^2$}) in above-ground vegetation, 
below-ground vegetation (roots), and soil pools, respectively. These states evolve according to the system of coupled 
ordinary differential equations
\begin{align}
\dstateV(\Time) &= \alphaV \NPP(\stateV(\Time), \forcing(\Time)) - \frac{\stateV(\Time)}{\tauV} \\
\dstateR(\Time) &= (1.0 - \alphaV) \NPP(\stateV(\Time), \forcing(\Time)) - \frac{\stateR(\Time)}{\tauR} \nonumber \\ 
\dstateS(\Time) &= \frac{\stateR(\Time)}{\tauR} + \frac{\stateV(\Time)}{\tauV} - \frac{\stateS(\Time)}{\tauS}, \nonumber
\end{align}
where the model driver $\forcing(\Time)$ is given by photosynthetically active radiation 
(\textrm{MJ/$m^2$/day}), and the dynamics rely on the following parameterized model of 
Net Primary Productivity (NPP; \textrm{kg C/$m^2$/day})
\begin{align}
\NPP(\stateV, \forcing) &= (1 - \fracRespiration) \GPP(\stateV, \forcing) \\
\GPP(\stateV, \forcing) &= \forcing \cdot \LUE \cdot \left[1 - \exp\left\{-\KEXT \cdot \LAI(\stateV) \right\} \right] \nonumber \\
\LAI(\stateV) &= \LAR \cdot \stateV, \nonumber
\end{align} 
where $\GPP(\stateV, \forcing)$ and $\LAI(\stateV)$ model Gross Primary Productivity (GPP; \textrm{kg C/$m^2$/day})
and Leaf Area Index (LAI; \textrm{$m^2/m^2$}), respectively.
Given a noisy synthetic model driver time series and initial conditions $\{\vegInit, \rootInit, \soilInit\}$,
we numerically solve the ODE at a daily time step
via the basic Euler scheme as implemented in the R \verb+BayesianTools+ package \citep{vsem}. 

\bibliography{surrogates} 

\end{document}

%% file: macros_general.tex
\usepackage{comment}
\usepackage{xcolor}

\usepackage{todonotes}

\newcommand*{\abs}[1]{\left\lvert#1\right\rvert}
\newcommand{\R}{\mathbb{R}}

\newcommand{\Exp}[1]{\exp\mathopen{}\left\{#1\right\}\mathclose{}}

\newcommand{\Def}{\coloneqq} % coloneqq comes from the mathtools package. 
\let\oldd\d\renewcommand\d{\relax\ifmmode\mathrm{d}\else\oldd\fi} %make \d be mathrm{d} in math mode, usual defn as underdot(?) in text mode
\newcommand{\proptoAdd}{\overset{\mathrm{add}}{\propto}} % Absorbs additive constants

\newcommand{\indicator}[1]{\mathds{1}_{#1}}

\DeclareMathOperator*{\argmin}{argmin}
\DeclarePairedDelimiterX\innerp[2]{(}{)}{#1\delimsize\vert\mathopen{}#2}

\newcommand*{\norm}[1]{\left\lVert#1\right\rVert}

\newcommand{\E}{\mathbb{E}}

\newcommand{\Cov}{\mathrm{Cov}}

\newcommand{\Gaussian}{\mathcal{N}}
\newcommand{\LN}{\mathrm{LN}} % Log-normal distribution 
\newcommand{\given}{\mid} 
\DeclarePairedDelimiterX{\divergencex}[2]{(}{)}{%
  #1\;\delimsize\|\;#2%
}

\newcommand{\KL}{\mathcal{D}_{\mathrm{KL}}\divergencex}

%% file: macros.tex
\usepackage{dsfont}

\newcommand{\Par}{u}
\newcommand{\parSpace}{\mathbb{U}} % Parameter space
\newcommand{\dimPar}{D} % Parameter dimension 
\newcommand{\obs}{y} % Data observation (response) vector
\newcommand{\noise}{\epsilon} % Random variable representing noise, typically in additive noise model. 
\newcommand{\obsSpace}{\mathbb{Y}} % Output space
\newcommand{\dimObs}{P} % Dimension of data observation vector
\newcommand{\fwd}{\mathcal{G}} % Forward model
\newcommand{\lik}{\mathsf{L}} % Likelihood 
\newcommand{\priorDens}{\pi_0} % Prior density 
\newcommand{\postDens}{\pi} % Unnormalized posterior density.
\newcommand{\postDensNorm}{\overline{\pi}} % Normalized posterior density.
\newcommand{\normCst}{Z} % Normalizing constant for posterior density. 
\newcommand{\likPar}{\Sigma} % Likelihood parameter

\newcommand{\Em}[1]{{#1}_{\star}}
\newcommand{\emMean}{\Em{\mu}}
\newcommand{\emSD}{\Em{s}}
\newcommand{\emVar}{\emSD^2}

\newcommand{\target}{\mathsf{f}} % Target map for emulation
\newcommand{\targetEm}{\Em{f}} % Emulator for target quantity
\newcommand{\targetTraj}{f} % Trajectory of emulator for target map
\newcommand{\emDist}{\nu} % Predictive distribution (law) of the emulator (a stochastic process)
\newcommand{\emDistPrior}{\emDist_0} % Prior predictive distribution of emulator
\newcommand{\emE}{\E_{\emDist}} % Expectation wrt emulator distribution.
\newcommand{\emObs}{z} % The design set used to train the emulator.
\newcommand{\targetRange}{\mathbb{F}} % Codomain for the target function.
\newcommand{\emLik}{\lik_{\target}} % Likelihood for emulator model for training on simulation data

\newcommand{\Ndesign}{N}
\newcommand{\GP}{\mathcal{GP}} % GP distribution.
\newcommand{\gpMeanBase}{\mu} % The base notation used for GP mean (no sub/superscripts). 
\newcommand{\gpKerBase}{k}

\newcommand{\gpMean}[1][\star]{\gpMeanBase_{#1}}
\newcommand{\gpKer}[1][\star]{\gpKerBase_{#1}}

\newcommand{\ep}{\mathrm{ep}}
\newcommand{\eup}{\mathrm{eup}}
\newcommand{\postEm}{\Em{\postDens}}
\newcommand{\postNormEm}{\Em{\postDensNorm}}
\newcommand{\normCstEm}{\Em{\normCst}}
\newcommand{\postApproxEP}{\Em{\postDensNorm}^{\ep}}
\newcommand{\postApproxMean}{\Em{\postDens}^{\mathrm{mean}}}
\newcommand{\postApproxEUP}{\Em{\postDens}^{\eup}}
\newcommand{\postApproxEUPNorm}{\Em{\postDensNorm}^{\eup}}
\newcommand{\postApproxNormMean}{\Em{\postDensNorm}^{\mathrm{mean}}}

\newcommand{\likApproxEP}{\Em{\lik}^{\ep}}
\newcommand{\likApproxEUP}{\Em{\lik}^{\eup}}

\newcommand{\qMeas}{Q} % Generic prob measure on parSpace (used in variational formulation)
\newcommand{\qDens}{q} % Density of qDens.
\newcommand{\qSpace}{\mathcal{Q}} % Space of densities or measures.
\newcommand{\loss}{\mathcal{L}}
\newcommand{\qDensOpt}{\qDens_{\mathrm{opt}}} % Solution of optimization problem
\newcommand{\qMeasOpt}{\qMeas_{\mathrm{opt}}}

\newcommand{\jgap}{\Delta_\normCst} % The "Jensen gap" for the normalizing constant.

\newcommand{\jointKOH}{\zeta} % Joint distribution over both surrogate and calibration parameters (as in KOH setup)
\newcommand{\postKOH}{\jointKOH^{\obs, \emObs}} % Posterior of the joint distribution
\newcommand{\condMargKOH}{\jointKOH^{\obs,\targetTraj}_{\Par}} % Marginal conditional of \Par, given \targetTraj and \obs.
\newcommand{\qCond}{T_{\qMeas}} % In the KOH context we have \qMeas(d\Par,d\targetTraj) = \emDist(d\targetTraj)\qCond(\targetTraj, d\Par)
\newcommand{\qSpaceCut}{\qSpace_{\mathrm{cut}}}

\newcommand{\emJoint}{\eta} % Joint measure over (u,f)
\newcommand{\BorelSig}{\mathcal{B}}
\newcommand{\emSpace}{\mathcal{F}}
\newcommand{\emSig}{\mathcal{A}} % Sigma algebra in function space for emulator
\newcommand{\refMeas}{\lambda} % Reference measure on parameter space

\newcommand{\sampleIndex}{k} % Generic index for samples. 
\newcommand{\NSample}{K} % Generic variable for algorithms that draw a set of samples (e.g., MCMC)

\newcommand{\accProbMH}{\alpha} % Metropolis-Hastings acceptance probability
\newcommand{\nextstate}[1]{{#1}^\dagger}

\newcommand{\propDist}{Q} % Proposal distribution (measure)
\newcommand{\propDistPar}{\propDist_{\Par}}

\newcommand{\propDistTarget}{\propDist_{\target}}
\newcommand{\propPar}{\tilde{\Par}} % Proposed parameter value
\newcommand{\pcnCor}{\rho}

\newcommand{\Time}{t}
\newcommand{\state}{x}
\newcommand{\odeRHS}{F}
\newcommand{\timeStart}{\Time_0}
\newcommand{\timeEnd}{\Time_1}
\newcommand{\stateIC}{\state_{\circ}}

\newcommand{\solutionOp}{\mathcal{S}}
\newcommand{\obsOp}{\mathcal{H}}

\newcommand{\forcing}{w}

\newcommand{\fwdLin}{G} % Linear forward model
\newcommand{\priorMean}{m_0} 
\newcommand{\priorCov}{C_0}
\newcommand{\postMean}{m}
\newcommand{\postCov}{C}
\newcommand{\noiseCov}{\Sigma}
\newcommand{\emBias}{r}
\newcommand{\emCov}{Q}
\newcommand{\postMeanEP}{\postMean^{\ep}}
\newcommand{\postMeanEUP}{\postMean^{\eup}}
\newcommand{\postCovEP}{\postCov^{\ep}}
\newcommand{\postCovEUP}{\postCov^{\eup}}
\newcommand{\epGain}{H} % = \postCov \fwdLin^\top \noiseCov^{-1}
\newcommand{\eupNoiseCov}{\tilde{\noiseCov}} % = \noiseCov + \emCov
\newcommand{\eupObs}{\tilde{\obs}} % = \obs - \emBias

\newcommand{\stateV}{\state_{\textrm{v}}}
\newcommand{\stateR}{\state_{\textrm{r}}}
\newcommand{\stateS}{\state_{\textrm{s}}}
\newcommand{\dstateV}{\dot{\state}_{\textrm{v}}}
\newcommand{\dstateR}{\dot{\state}_{\textrm{r}}}
\newcommand{\dstateS}{\dot{\state}_{\textrm{s}}}
\newcommand{\NPP}{\textrm{NPP}}
\newcommand{\GPP}{\textrm{GPP}}
\newcommand{\alphaV}{\alpha_{\textrm{v}}}
\newcommand{\tauV}{\tau_{\textrm{v}}}
\newcommand{\tauR}{\tau_{\textrm{r}}}
\newcommand{\tauS}{\tau_{\textrm{s}}}
\newcommand{\LUE}{\ell}
\newcommand{\LAR}{r}
\newcommand{\KEXT}{\kappa}
\newcommand{\fracRespiration}{\gamma}
\newcommand{\LAI}{\textrm{LAI}}
\newcommand{\vegInit}{\stateV^0}
\newcommand{\rootInit}{\stateR^0}
\newcommand{\soilInit}{\stateS^0}

\newcommand{\perm}{\kappa}
\newcommand{\pressure}{v}
\newcommand{\source}{s}
\newcommand{\stateObs}{\state^{\mathrm{obs}}}